\input harvmac
\def\R{{\bf R}}
\def\CM{{\cal M}}

\def\K3{{\bf K3}}
\def\journal#1&#2(#3){\unskip, \sl #1\ \bf #2 \rm(19#3) }
\def\andjournal#1&#2(#3){\sl #1~\bf #2 \rm (19#3) }

\def\tilde{\widetilde}

\def\frac#1#2{{#1\over#2}}

\def\half{\frac12}

\def\inbar{\,\vrule height1.5ex width.4pt depth0pt}
\def\IC{\relax\hbox{$\inbar\kern-.3em{\rm C}$}}
\def\IR{\relax{\rm I\kern-.18em R}}
\def\IP{\relax{\rm I\kern-.18em P}}
\def\Z{{\bf Z}}

%
%

%
\catcode`\@=11
\def\slash#1{\mathord{\mathpalette\c@ncel{#1}}}
\overfullrule=0pt

\def\KK{{\cal K}}

\def\NN{{\cal N}}

\def\underrel#1\over#2{\mathrel{\mathop{\kern\z@#1}\limits_{#2}}}

\catcode`\@=12


%

\def\det{{\rm det}}

\def \sinh{{\rm sinh}}

\def\det{{\rm det}}
\def\exp{{\rm exp}}


\def\bS{{\bf S}}

\def\bT{{\bf T}}
\def\bX{{\bf X}}


\rightline{IASSNS-HEP-99-27}
\Title{
\rightline{hep-th/9903224}
} {\vbox{\centerline{The D1/D5 System and Singular CFT}}}
\medskip

\centerline{\it Nathan Seiberg and Edward Witten}
\bigskip
\centerline{School of Natural Sciences}
\centerline{Institute for Advanced Study}
\centerline{Olden Lane, Princeton, NJ 08540}

\smallskip

\vglue .3cm
\bigskip
\noindent
We study the conformal field theory of the D1/D5 system compactified
on $\bX$ ($\bX$ is $\bT^4$ or $\K3$).  It is described by a sigma
model whose target space is the moduli space of instantons on $\bX$.
For values of the parameters where the branes can separate, the
spectrum of dimensions in the conformal field theory exhibits a
continuum above a gap.  This continuum leads to a pathology of the
conformal field theory, which explains a variety of problems in various
systems.  In particular, we explain the apparent discrepancy between
different methods of finding the spectrum of chiral fields at certain
points in the moduli space of the system.

\noindent
\Date{2/99}
\def\AdS{{\bf AdS}}
\def\S{{\bf S}}

\newsec{Introduction}

The D1/D5 system, which has been much studied of late,
is constructed from branes in $\R^6\times \bX$, where
$\bX$ is $\bT^4$ or $\K3$.  One considers $Q_5$ D5-branes
wrapped on $X$, making strings in $\R^6$; and  one adds $Q_1$
D1-branes that are localized on $\bX$ and parallel to the strings
made from the fivebranes.  If $Q_1$ and $Q_5$ are large,
this system has a supergravity description as a ``black string,''
whose near horizon geometry \ref\nearho{G. Gibbons and P. Townsend,
``Vacuum Interpolation In Supergravity Via Super $p$-Branes,''
Phys. Rev. Lett. {\bf 71} (1993) 5223; G. W. Gibbons and P. Townsend,
``Higher Dimensional Resolution Of Dilatonic Black Hole Singularities,''
Class. Quant. Grav. {\bf 12} (1995) 297.}
looks like $\AdS_3\times \S^3\times \bX$.
In this limit, the D1/D5 system
 is believed \ref\malda{J. Maldacena, ``The Large $N$ Limit Of
 Superconformal Field Theories And Supergravity,''
Adv. Theor. Math. Phys. {\bf 2} (1997) 231,  hep-th/9711200.}
 to be described by a conformal
field theory on the boundary of $\AdS_3\times \S^3$.  (The boundary
in question is a conformal boundary in the sense of Penrose
\ref\penrose{R. Penrose and W. Rindler, {\it Spinors And Spacetime},
vol. 2 (Cambridge University Press, 1986), chapter 9.}.)

One can generalize the D1/D5 system by turning on various other
fields or charges, such as theta angles, RR fields,
 or D3-branes wrapped on two-cycles
in $\bX$.  However, the simplest 
D1/D5 system with no such extra fields 
has the basic property that the branes can separate at no cost in energy.
In fact, any collection of D5 branes (wrapped on $\bX$) and D1 branes
(localized on $\bX$), with all branes parallel in $\R^6$, is BPS-saturated
(if the theta angles and RR fields are zero).
Hence, the D1/D5 system can break into pieces at no cost in energy.
The goal of the present paper is to study the implications of this
for the boundary conformal field theory.

One description of the low energy physics of this system is provided
by the $U(Q_5)$ gauge theory on the D5-branes.  The D1-branes can be
interpreted as $Q_1$ instantons in this gauge theory, and the
D3-branes correspond to magnetic fluxes, representing a non-zero first
Chern class.  We denote by $\CM$ the moduli space of $U(Q_5)$
instantons on $\bX$ with onebrane charge\foot{For $\bX={\bf T}^4$, the
onebrane charge $Q_1$ equals the instanton number $Q_1'$, while for
$\bX={\bf K3}$, $Q_1=Q_1'-Q_5$.  We will loosely refer to $Q_1$ as the
instanton number.} $Q_1$ and with $c_1$ determined by the number of
threebranes.  The dynamics of our system is described by a $(4,4)$
sigma model whose target space is $\CM$.

To reproduce the ``pure'' D1/D5 system, we set $c_1=0$.  Then we expect
to see in the gauge theory that the branes can separate.  In fact,
an instanton can become ``small,'' and separate from the D1/D5 system
as a D1-brane.  Or the structure group of a $U(Q_5)$ instanton
might reduce to $U(Q_5')\times U(Q_5'')$
with instanton numbers $Q_1'$ and $Q_1''$ in the two factors 
(and $Q_5'+Q_5''=Q_5$, $Q_1'+Q_1''=Q_1$).  Then the two groups of
branes, with respective quantum numbers $(Q_1',Q_5')$ and $(Q_1'',Q_5'')$,
can separate in $\R^6$.  A special case of this, with $(Q_1',Q_5')=
(0,1)$, is the emission of a fivebrane, a process that is $T$-dual to
shrinking of an instanton and its emission as a D1-brane.

In any of these cases, the separation of the branes is described in terms of
gauge theory by a passage from a Higgs to a Coulomb branch.
To see this, one uses an effective description as a two-dimensional
gauge theory on the string.  For example, the emission of a small instanton
is described by an effective $U(1)$ gauge theory in $1+1$ dimensions
with $Q_5$ hypermultiplets of charge 1.  This theory has a Higgs branch,
describing a non-small instanton, and a Coulomb branch, describing
a D1-brane that has been ejected from the D1/D5 system.  The two
branches meet at the small instanton singularity.  Similarly,
the splitting of the $(Q_1,Q_5)$ system to $(Q_1',Q_5')$ plus
$(Q_1'',Q_5'')$ 
means that the structure group of the instanton reduces from
$U(Q_5)$ to $U(Q_5')\times U(Q_5'')$, leaving unbroken an extra $U(1)$.
This $U(1)$ is coupled  to several charged hypermultiplets
(which represent the instanton moduli
 that must be set to zero to reduce the structure
group of the instanton to $U(Q_5')\times U(Q_5'')$).  The low energy
theory describing the splitting is $U(1)$ coupled to these hypermultiplets.

In supersymmetric gauge theories above two dimensions, the Higgs and
Coulomb branches parametrize families of supersymmetric vacua.  
In two dimensions, this is not so; the usual infrared divergences
of massless bosons cause the quantum wave functions to spread out over
the Higgs or Coulomb branches.  Nevertheless, even in two dimensions,
the Higgs and Coulomb branches are described by different conformal
field theories \ref\witcom{E. Witten, ``Some Comments On String
Dynamics,'' hep-th/9507121,
in Strings 95, {\it Future Perspectives In String Theory}, ed. I. Bars 
et. al.}.  The different branches have different $R$-symmetries and usually
have different central charges.

So even though the two branches meet classically, they are
disconnected in conformal field theory.  How can this happen?  One
idea is that as one flows to the infrared, the metric on the two
branches might be renormalized 
\ref\dps{M. Douglas, J. Polchinski, and A. Strominger, ``Probing
Five-Dimensional Black Holes WIth $D$-Branes,'' hep-th/9703031, JHEP 
{\bf 9712} (1997) 3.}
so that the classical meeting place of
the two branches would be ``infinitely far away'' as seen on either
branch.  Something like this happens to the Coulomb branch at the
one-loop level for any values of $Q_1$ and $Q_5$ for which both a
Coulomb and Higgs branch exist
\ref\seidai{E. Diaconescu and N. Seiberg, ``The Coulomb Branch Of
$(4,4)$ Supersymmetric Field Theories In Two Dimensions,''
hep-th/9707158, JHEP {\bf 9707} (1997) 1.}.
For a single vector multiplet, the classical Coulomb branch is a copy
of $\R^4$; we regard the point of intersection with the classical
Higgs branch as the origin of $\R^4$ and let $u$ denote a radial variable
on $\R^4 $ that vanishes at the origin.
The classical metric  $du^2+u^2d\Omega^2$ 
of the
Coulomb branch  is renormalized at the one-loop level so that near
$u=0$ it looks like
\eqn\ujjcu{{1\over u^2}\left(du^2+u^2d\Omega^2\right).}
Thus $u=0$ is at infinite  distance.  Moreover, there is a ``Liouville
coupling'' $R\ln u$ (with $R$ the worldsheet scalar curvature) such
that the string coupling constant diverges as one approaches $u=0$.
The metric \ujjcu\ describes an infinitely long tube near $u=0$, which
is why we speak of ``tubelike''  behavior, and the string coupling
diverges as one goes down the tube.

One might hope for something similar on the Higgs branch.
But the fate of the Higgs branch must be more subtle.  One cannot
simply find a quantum correction generating a tubelike metric, since
gauge loops do not renormalize the hyper-K\"ahler metric on the Higgs branch
\ref\aps{P.C.~Argyres, M.~Ronen Plesser and N.~Seiberg,
``The Moduli Space of Vacua of N=2 SUSY QCD and Duality in N=1 SUSY
QCD,'' Nucl. Phys. {\bf B471} (1996) 159, hep-th/9603042.}.
However, the description of the Higgs branch via a sigma model with
target space the classical Higgs branch ${\cal M}$ is not good near
the singularities of ${\cal M}$.  One might hope that in terms of some
new variables that do give an effective description near the singularity
one would find a tubelike behavior.
\nref\vafoog{H. Ooguri and C. Vafa, ``Two-Dimensional Black Hole
And Singularities Of CY Manifolds,'' Nucl. Phys. {\bf B463} (1996) 55.}
\nref\ghm{R. Gregory, J. A. Harvey, and G. Moore, ``Unwinding Strings
And $T$-Duality Of Kaluza-Klein And $H$-Monopoles,'' Adv. Theor.
Math. Phys. {\bf 1} (1998) 283.}%
This is known in certain special
cases via duality between Type IIA at an ${\bf A}_{n-1}$ 
singularity and Type IIB with $n$ parallel D5-branes \refs{\vafoog,\ghm}.
In more generality, such a description can be obtained using
duality between the Higgs and Coulomb branches of $(4,4)$ supersymmetric
gauge theories in two dimensions \nref\brodie{J. Brodie, ``Two-Dimensional
Mirror Symmmetry From $M$-Theory,'' hep-th/9709228.}%
\nref\sethi{S. Sethi,
``The Matrix Formulation of Type IIb Five-branes,''
Nucl. Phys. {\bf B523} (1998) 158,
hep-th/9710005.}%
\refs{\brodie,\sethi}.

The purpose of the present paper is to exhibit tubelike behavior near the
singularities of the D1/D5 system, and therefore various other systems
to which it is dual.  We do this by elaborating upon a construction
introduced by Maldacena, Michelson, and Strominger \ref\mms{
J. Maldacena, J. Michelson, and A. Strominger, ``Anti-de Sitter
Fragmentation,'' hep-th/9812073.}.  
We will give a description of the physics of
the Higgs branch near its singularity in terms of an effective two-dimensional
field $\phi$.  $\phi$ will be a Liouville field with kinetic energy $|d\phi|^2$
and a Liouville coupling $\phi R$.  The classical singularity of the Higgs
branch, or in other words its meeting with the classical Coulomb branch,
corresponds to $\phi=\infty$.  Because of the Liouville term, the string
coupling blows up as one goes to $\phi=\infty$.

Thus, whether one begins on the Coulomb branch or the Higgs branch, the
meeting place of these two branches is at infinite distance in terms of
the right variables.  Hence, starting from either branch, one can never
reach the other.  Starting from either branch and trying to approach
the second, one must go ``down the tube'' and the string coupling constant
blows up.  The blowup of the string coupling constant is extremely
important in some applications of the systems that have this behavior
(like Type IIA at an ${\bf A}_{n-1}$ singularity) because it makes
it possible to have nonperturbative effects (in that case, enhanced
gauge symmetry) that cannot be turned off by going to weak coupling. 

Since the usual D1/D5 system is the quantum mechanics of a Higgs
branch, the tubelike nature of the singularity of the Higgs branch
has specific consequences for the boundary conformal field theory
that governs this system.    
Liouville theory has a continuous spectrum of dimensions above a certain
threshold.  So the boundary conformal field theory will have this property.
Also, in Liouville
theory with a strong coupling end that is not cut off or protected
in any way, correlation functions generally diverge from integration
over the Liouville field.   So correlation functions of the D1/D5
conformal field theory can be expected to receive divergent contributions
near the singularity.

Using S duality, we can transform the problem to $Q_5$ NS5-branes and
$Q_1$ fundamental strings
\ref\msex{J.~Maldacena and A.~Strominger,
``AdS(3) black holes and a stringy exclusion principle,''
JHEP {\bf 12} (1998) 005, hep-th/9804085.}.  
In this case, one can study the system by using conformal field theory
of the fundamental strings, and one can also hope to compare this to
the boundary conformal field theory at infinity.  However, such a 
comparison will be affected by the continuous spectrum of dimensions
mentioned in the last paragraph.  For example, above the threshold,
chiral primary states lie in the continuum, and one should expect
difficulty in counting them.

In this discussion, we have emphasized the $\AdS_3$ examples.  But
part of our discussion is also relevant to $\AdS_n$ for $n>3$.

The discussion in the present paper is reminiscent of
``the membrane at the end of the universe'' 
\ref\duff{M. Duff and C. Sutton,
``The Membrane At The End Of The Universe,'' New. Sci. {\bf 118} 
(1988) 67.}
and the Liouville theory of $\AdS_3$
\ref\CHvD{O.~Coussaert, M.~Henneaux and P.~van Driel, ``The Asymptotic
dynamics of three-dimensional Einstein gravity with a negative
cosmological constant,'' Class. Quant. Grav. {\bf 12} (1995) 2961.},
though the interpretation seems to be somewhat different.

This paper is organized as follows.
In section 2, we review the moduli space of the
D1/D5 system, explaining more precisely what conditions on the moduli
are needed to get the singularity that we will be studying.

In section 3, we show how the study of large strings or branes introduced
in \mms\ gives an effective description of the singularity of the Higgs
branch.  We also show how, for $\AdS_n$ models with $n>3$, one reproduces
in this way expected properties of the boundary conformal field theories.
Among other things, we show that the boundary must have positive scalar
curvature for stability, in agreement with the known behavior of $N=4$
super Yang-Mills theory in four dimensions.

In section 4, we analize more quantitatively the effective theory of
the long string in $\AdS_3$, and we determine the exact value of the
threshold above which the conformal field theory has a continuous
spectrum.
 
In section 5, we discuss the fate of the chiral states,
and some additional applications.

\def\T{{\bf T}} 
\def\Z{{\bf Z}}
\def\K{{\cal K}}

\newsec{Near Horizon Moduli Space And Singularities}

In this section, we will review the moduli space of $\AdS_3$
compactifications, following \ref\dijk{R. Dijkgraaf, ``Instanton
Strings And Hyperk\"ahler Geometry,'' hep-th/9810210.},
and then describe precisely where a singularity is expected. 

\def\KK3{{\bf K3}}
\subsec{Classification Of Models}

Consider string compactification on $\bX=\T^4$ or ${\bf K3}$.
The duality group is $ \K= SO(5,n;\Z)$, where $n=5$ for $\T^4$ and $n=21$ for
$\KK3$.  
The moduli space of vacua is
\eqn\uxu{{\cal W}=SO(5,n;\Z)\backslash SO(5,n;\R)/SO(5)\times SO(n).}

We now want to consider a system consisting of parallel strings
in $\R^6\times \bX$.  The D1/D5 system is the special case in which
the strings are either D1-branes or else D5-branes wrapped on $\bX$.
In general, the charge of a string is measured by a charge vector $v$
that takes values in an even, unimodular lattice $\Gamma^{5,n}$ on which
$\K$ acts.  The quadratic form on $\Gamma^{5,n}$ has five positive
and $n$ negative eigenvalues.   $\Gamma^{5,n}$ can be embedded
in a vector space $V$ which has metric of signature $(5,n)$. 
A point in ${\cal W}$ 
gives a decomposition $V=V_+\oplus V_-$, where the quadratic
form is positive on $V_+$ and negative on $V_-$.  $\Gamma^{5,n}$ is
analogous to a Narain lattice in string theory, and $V_\pm$ are analogous
to the spaces of right-moving and left-moving momenta.

For a string with charge vector $v$ to be a BPS configuration, it is necessary
to have $v^2\geq 0$.  Since the lattice is even, we have
\eqn\knxon{v^2=2N,~~{\rm with}~N\geq 0.}

Given any vector $v$ and any point in moduli space, we write
$v=v_++v_-$, with $v_\pm \in V_\pm$.  The tension of a string with
charge $v$ is then (up to a multiplicative
constant independent of $v$ and the moduli)
\eqn\hxi{T(v)=|v_+|.}

We are mainly interested in the case that the charge vector $v$ is
``primitive,'' in other words is
not of the form $v=kv'$ with $v'$ a lattice vector
and $k$ an integer greater than 1.  The reason for this restriction
is that otherwise the model is singular -- capable of breaking into
subsystems at no cost in energy -- for all values of the moduli.

Now, any two primitive 
lattice vectors $v$ and $w$ with $v^2=w^2=2N$ are equivalent
up to a transformation by an element of $\K$, as explained in \dijk.
So up to a duality transformation, there is only one model
for every positive integer $N$.  
$N=0$ is the case of a single elementary string, so we are really
interested in $N>0$.

For example, the D1/D5 system is the case that there is a decomposition
\eqn\koko{
\Gamma^{5,n}=\Gamma^{1,1}\oplus \Gamma^{4,n-1},} where $\Gamma^{1,1}$
is a two-dimensional sublattice whose quadratic form in a suitable basis
looks like
\eqn\jjxn{\left(\matrix{ 0 & 1 \cr 1 & 0 \cr}\right),}
and the moduli are such that the decomposition \koko\ commutes with
the projection to $V_+$ and $V_-$.  If the theta angles and RR fields
vanish,
then (as one can see in more detail from formulas given in \dijk)
there is such a decomposition with
the D1 string and the string built from a wrapped D5 brane represented by
null vectors $(1, 0)$ and $(0,1)$ in $\Gamma^{1,1}$.  The
D1/D5 system can then be described by
 charges $(Q_1,Q_5)\in \Gamma^{1,1}$, and hence 
\eqn\jurry{N=Q_1Q_5.}
In this construction,
$v$ being primitive means that $Q_1$ and $Q_5$ are relatively prime.

As we have discussed in the introduction, the D1/D5 system is expected
to have, for each value of $Q_1$ and $Q_5$, a singularity associated
with a small $U(Q_5)$ instanton.  To achieve such a singularity, one
must adjust $4(Q_5-1)$ parameters in the instanton solution.  The
number of parameters that must be adjusted depends on $Q_5$, so the
small instanton singularity of the D1/D5 system depends on the value
of $Q_5$, and not on the product $N=Q_1Q_5$.  Thus, a single system,
characterized by the choice of one integer $N$, has different
singularities corresponding to the different factorizations $N=Q_1Q_5$
with $Q_1$ and $Q_5$ relatively prime.  To clarify this further, we
will in section 2.2 analyze precisely where there are singularities of
the near horizon theory.

\bigskip
\noindent{\it Near Horizon Moduli Space}

Given a choice of charge vector $v$, one can construct a supergravity
solution that describes a string of that charge.  Here one finds
\ref\fixedscalars{ S. Ferrara, R. Kallosh, and A. Strominger, ``$N=2$
Extremal Black Holes,'' Phys. Rev. {\bf D52} (1995) 5412,
hep-th/9508072; S. Ferrara, G. W. Gibbons, and R. Kallosh, ``Black
Holes And Critical Points In Moduli Space,'' Nucl. Phys. {\bf 500}
(1997) 75, hep-th/9702103.} an interesting phenomenon: in the field of
this string, the moduli are not constant, and vary in such a way that
at the horizon, the vector $v$ is ``purely right-moving,'' that is, it
lies in $V_+$.  The moduli that would rotate $V_+$ and $V_-$ so that
$v$ no longer lies in $V_+$ are ``fixed scalars''; in the near horizon
geometry of the string, they are massive.

\def\H{{\cal H}}
The near horizon geometry therefore has a reduced moduli space.  Roughly
it is a Narain moduli space of signature $(4,n)$, since $v$ is now
constrained to lie in $V_+$ and only the four-dimensional orthocomplement
of $v$ in $V_+$ is free to vary.  The near horizon geometry
also has a reduced duality group, namely the subgroup $\H$
of $\K$ consisting of transformations that leave fixed the vector $v$.
The moduli space of the near horizon geometry is
\eqn\kiffo{{\cal N}=\H\backslash SO(4,n;\R)/SO(4)\times SO(n).}

\subsec{Location Of Singularities}

  Now we come to the question of under
what condition the conformal field theory that describes the near horizon
physics becomes singular.  As we explained in the introduction, this should
occur when the charge  vector $v$ can be written as $v=v'+v''$,
where $v'$, $v''$, and $v$ are all mutually BPS.
In view of \hxi, the condition for this is that $|v_+|=|v'_+|+|v''_+|$.
Since $v=v_+$ in the near horizon geometry, this is $|v|=|v'_+|+|v''_+|$.
In view of the triangle inequality, this is equivalent to saying that
the projection of $v'$ (or of $v''$) to $V_+$ is a multiple of $v$.
In other words, the lattice $\Gamma'$ generated by $v$ and $v'$ has
a projection to $V_+$ that lies in a one-dimensional subspace, the space
of multiples of $v$.  

Assuming that this is so, the projection of $\Gamma'$ to $V_+$ is
one-dimensional, being generated by $v$, so $\Gamma'$ has signature $(1,1)$.
Given any primitive $v'\in \Gamma'$ and not a multiple of $v$, 
the requirement that the projection of $v'$ to $V_+$ is a multiple
of $v$ puts four conditions on the near horizon moduli.  ($V_+$ is
five dimensional, so asking that a vector in $V_+$ be a multiple of
$v$ imposes four conditions.)  If these conditions are imposed, then
(as $v$ and $v'$ generate $\Gamma'$) 
the projection of $\Gamma'$ to
$V_+$ consists of multiples of $v$.
When this happens, every way of writing $v=v'+v''$ with $v',v''\in \Gamma'$
and 
$(v')^2,\,(v'')^2\geq 0$ will be a way of breaking our system into
two BPS subsystems at no cost in energy.  The ability to do this should
give a singularity in the boundary conformal field theory.

Thus the loci in moduli space on which the CFT is expected to be
singular are classified by signature $(1,1)$ sublattices $\Gamma'\subset 
\Gamma$ that
contain $v$.\foot{Of course, we are only interested
in the choice of $\Gamma'$ up to the action of the duality group $\H$ that
keeps $v$ fixed.  Equivalently, we want to choose the pair $\Gamma'$ and $v$
up to the action of the full duality group $\K$.} 
For each such $\Gamma'$, a singularity is found
by adjusting one hypermultiplet so that all projections of vectors
in $\Gamma'$ to $V_+$ become proportional.  On this locus, the
string can break up according to any decomposition $v=v'+v''$
with $v',v''\in\Gamma'$ and
$(v')^2,\,(v'')^2\geq 0$.  (For some $\Gamma'$, there may exist no such
$v'$, $v''$, and then there is no singularity associated with $\Gamma'$.)

For example, let us classify all cases in which $\Gamma'$ is unimodular,
in other words in some basis its quadratic form is
\eqn\umbo{\left(\matrix{ 0 & 1\cr 1 & 0 \cr}\right).}
To put it differently, $\Gamma'$ is isomorphic to the unique even unimodular
lattice $\Gamma^{1,1}$ of signature $(1,1)$.
Such a $\Gamma'$ has up to a duality transformation
a unique embedding in $\Gamma^{5,n}$ (this is proved by noting that the
orthocomplement of $\Gamma'$ is even and unimodular and hence unique
up to isomorphism).
The transformation that puts $\Gamma'$ in this standard form
may rotate $v$ into an arbitrary form, modulo the symmetries of $\Gamma'$
plus the fact that $v$ is primitive and $v^2=2N$.  So in general
we have $v=(Q_1,Q_5)$ for some relatively prime
integers $Q_1$ and $Q_5$ with $Q_1Q_5=N$.  By a symmetry of
$\Gamma'$ (namely $(Q_1,Q_5)\to (-Q_1,-Q_5)$), we can assume that
$Q_1$ and $Q_5$ are both nonnegative.  The only remaining symmetry
of $\Gamma'$ is $(Q_1,Q_5)\leftrightarrow (Q_5,Q_1)$.  (For example, for
$\bX=\T^4$, this is a $T$-duality transformation on $\bX$.) So, for given
$N$, choices of a unimodular $(1,1)$ lattice $\Gamma'$ containing
$v$ are classified by the unordered relatively prime pairs $Q_1,\,Q_5$
with $N=Q_1Q_5$.  

In other words, the singularities with unimodular $\Gamma'$ are
the small instanton and partial un-Higgsing singularities of
$U(Q_5)$ gauge theory described in the introduction, 
for various values of $Q_5$.
There are also singularities for which  $\Gamma'$ is not unimodular.
(For example, if $v$ describes the D1/D5 system while
$v'$ has threebrane charge as well as D1 and D5 charge,
then $v$ and $v'$ can generate a lattice that is not unimodular.)
In the present paper, our general analysis of the tube behavior via
long strings in section 3 is valid for all of the singularities.
But our more quantitative study in section 4 uses specifically
the NS1/NS5 system (which is dual to D1/D5) and so is special to the
case of unimodular $\Gamma'$.

We pause here to point out a
subtlety that we have hidden in our exposition.  Let $\Gamma_\perp$ be
the sublattice of $\Gamma^{5,n}$ consisting of vectors perpendicular to $v$,
and let $\H'$ be the automorphism group of $\Gamma_\perp$.  $\H$ is
a subgroup of $\H'$ (since any symmetry of $\Gamma^{5,n}$ that leaves
$v$ fixed must map $\Gamma_\perp$ to itself), and is actually a proper
subgroup (any element of $\H'$ is a symmetry of the sublattice
$v\Z\oplus \Gamma_\perp$ of $\Gamma^{5,n}$, but may not extend to a symmetry
of $\Gamma^{5,n}$ itself).  To describe the singularities
associated with unimodular $\Gamma'$,
 we have in the text classified, up to the action of $\H$,
the  embeddings of $\Gamma^{1,1}$ in $\Gamma^{5,n}$
that contain $v$.  It would be tempting to reason as follows:
 Every such $\Gamma^{1,1}$ embedding contains a (unique up to sign) primitive
vector $w\in \Gamma_\perp$, with $w^2=-2N$.  The choice of $w$ determines
the lattice $\Gamma^{1,1}$. So why not classify the $\Gamma^{1,1}$'s by
classifying $w$ up to the action of $\H'$?   This reasoning actually gives the
wrong result ($w$ is unique up to the $\H'$ action, so one would
conclude that the singularity depends only on $N$ and not on $Q_5$), 
which is possible since $\H$, not $\H'$, is the symmetry group of the problem.

\bigskip\noindent{\it Examples}

We conclude this section by briefly stating some examples.
The properties we assert can all be verified in detail using formulas
in \dijk.

For the D1/D5 system, the fixed scalars, which are entirely absent
in the near horizon physics, are the volume of $\bX$, the
anti-self-dual part of the NS $B$-fields, and a linear combination
of the RR zero-form and four-form.  For example, in terms of the
description of the D1/D5 system by a sigma model with instanton moduli
space as the target, it is natural that the volume of $\bX$ should
drop out as the instanton equation on $\bX$ is conformally invariant.

In order to see a singularity from separating the D1/D5 system into
D1/D5 subsystems, four more parameters must be set to zero.  They are
the self-dual part of the NS $B$-fields and a linear combination of
the RR zero-form and four-form.  For example, it has been argued
\ref\noncom{N. Nekrasov and A. S. Schwarz, ``Instantons On Noncommutative
$\R^4$ and $(2,0)$ Superconformal Six-Dimensional Theory,'' hep-th/9802068,
Commun. Math. Phys. {\bf 198} (1998) 689.} that turning on the self-dual part 
of
the $B$-field deforms the instanton equations to the equations for
instantons in noncommutative Yang-Mills theory.  This operation eliminates
the small instanton singularity; it ``blows up'' the small instanton
locus by, in one description, adding a constant to the ADHM equations.
So this operation removes the singularity from breaking up the D1/D5
system into subsystems.

By an $S$-duality transformation, one can identify the corresponding
statements for the NS1/NS5 system.  The fixed scalars are the string
coupling constant, the anti-self-dual part of the RR $B$-fields, and a
linear combination of the RR zero-form and four-form.  There is a
singularity from breaking the NS1/NS5 system into similar subsystems
if the remaining RR fields -- the self-dual part of the two-form, and
a linear combination of the zero-form and four-form -- vanish.

\nref\gks{A.~Giveon, D.~Kutasov and N.~Seiberg, ``Comments on String
Theory on $AdS_3$,'' Adv. Theor. Math. Phys. {\bf 2} (1998) 733,
hep-th/9806194.}%
\nref\dbort{J.~de Boer, H.~Ooguri, H.~Robins and J.~Tannenhauser,
``String theory on AdS(3),'' hep-th/9812046.}%
\nref\ks{D.~Kutasov and N.~Seiberg, ``More Comments on String Theory
on $AdS_3$,'' EFI-99-8, IASSNS-HEP-99-30, to appear.}%

In particular, in the study of the NS1/NS5 system by conventional
conformal field theory methods \refs{\gks-\ks}, the RR fields are all
assumed to vanish.  Hence one is necessarily ``sitting'' on the
singularity.  As we explain in section 5, we believe that this is
responsible for some apparent discrepancies between computations
performed in the worldsheet and spacetime conformal field theories.

\bigskip\noindent{\it Comparison To Symmetric Product} 

We conclude this section with a brief discussion of the much-discussed
relation of
the spacetime conformal field theory of the near horizon system with
a given value of $N$ to a sigma  model with target space the symmetric
product of $N$ copies of $\bX$, which we denote as  $S^N\bX$.

For reasons explained in \refs{\gks, \dijk}, the sigma model with
$S^N\bX$ as target very likely has a moduli space that agrees with
\kiffo.  This alone suggests that the sigma model with target $S^N\bX$
is the right model.  Moreover, rigorous mathematical theorems\foot{For
an expository survey of mathematical results cited in this paragraph
with references, see
\ref\beauville{A. Beauville, ``Riemannian Holonomy
And Algebraic Geometry,'' math.AG/9902110.}.}
show that (for all $N$ and certain
charge vectors $v$ with $v^2=2N$; the theorem has not been proved for
all such vectors) the moduli space of instantons with suitably specified
Chern classes is indeed birational to a symmetric product of $N$ copies
of $\bX$.  Further, it has been proved recently 
that any two birational
compact hyper-K\"ahler manifolds are deformation equivalent.  We believe
that these facts mean that the D1/D5 system, for any $N$, is on the
moduli space of the symmetric product.

Nevertheless, the relation between them is extremely subtle.
  The hyper-K\"ahler
manifold $S^N\bX$ has singularities where two points meet.  Resolving
 the singularities and turning on the associated theta 
angle put us into a  moduli space that parametrizes unknown objects
that may have a rather complicated behavior.  If it is true that
this family of conformal field theories is the one we want, then these
conformal field theories exhibit quite an assortment of singularities
on many different loci in moduli space.  We cannot rule
out this hypothesis, and it seems plausible that it is true.  

For the usual
questions involving black holes in a macroscopic $\AdS$ model,
one wants large $Q_1$ and $Q_5$, leading to very special small instanton
singularities characteristic of the chosen charges.  This may be described
by a conformal field theory that can be connected to the symmetric
product, but it cannot be described by the symmetric product itself
-- which indeed depends only on $N$ and not on the separate choice of
$Q_1$ and $Q_5$, and so cannot possibly yield the right singularities.
We do not know where on the moduli space, if anywhere, the symmetric
product point might be.  

It is tempting to think that the symmetric
product point might be $Q_1=N$, $Q_5=1$, with vanishing theta angles
and RR fields.  But as in general the D1/D5 system has singularities
in codimension $4(Q_5-1)$, for $Q_5=1$ the system is ``generically singular''
whatever that means.  This does not sound like the hallmark of the
symmetric product point.  We make a few more remarks on $Q_5=1$ in
section 4.

\newsec{Mechanism For The Singularity}
 
In this section, by enlarging upon comments by Maldacena, Michelson,
and Strominger \mms,
we will give a microscopic mechanism for exhibiting
the singularity of the Higgs branch.  While the rest of the paper
focusses only on the $\AdS_3$ examples, in the present section we consider
also $\AdS_{D+1}$ for all $D\geq 2$.  ($D=1$ has special properties
explored in \mms.)   

As we explained in the introduction, the potential
singularity of the Higgs branch arises when
a brane is emitted from the system.  Emission of the brane will lower
the charges of the remaining system.  For example, suppose that
we are studying ${\cal N}=4$ super Yang-Mills in four dimensions,
with gauge group $SU(N)$ (or $U(N)$ if one takes into account a ``singleton''
degree of freedom at infinity), via
Type IIB on $\AdS_5\times \S^5$ with $N$ units of five-form flux on $\S^5$.
If the boundary of $\AdS_5$ were flat, it would be possible to move
on the Coulomb branch, Higgsing the $SU(N)$ down to $SU(N-1)\times U(1)$
by giving an expectation value to a scalar field $\phi$.
In such a vacuum, at very short distances $\ll 1/|\phi|$, one sees a
gauge group $SU(N)$, but at longer distances this is reduced to
$SU(N-1)\times U(1)$.  

Let us try to translate this mechanism into $\AdS$ using the familiar
IR-UV connection.  The fact that the gauge group in the conformal
field theory is $SU(N)$ at very short distances means that very near the
boundary of $\AdS_5$, there are $N$ units of five-form flux on $\S^5$.
But at longer distances in the conformal field theory, or in other words 
farther from the boundary of $\AdS_5$, there is only $SU(N-1)$, corresponding
to $N-1$ units of five-form flux.  But the flux in $\AdS_5$ can only
change in crossing a threebrane.  So we assume that there is a very
large region $W\subset \AdS_5$, with a threebrane wrapped on $\partial W$,
the boundary of $W$.  Then the five-form flux is $N-1$ in $W$ and $N$ outside.
If $W$ is very large, then $\partial W$ is
roughly speaking ``close to the boundary'' of $\AdS_5$.  We will call
a brane whose worldvolume is such a $\partial W$ a ``large brane,'' or, when
it is a one-brane, a ``long string.'' The $SU(N-1)\times 
U(1)$ low energy gauge symmetry of the Higgsed theory is in this situation
 interpreted as
a $U(1)$ carried by the large threebrane plus the $SU(N-1)$ described by
supergravity on $W\times \S^5$ with a flux of $N-1$.  

\nref\phirs{E. Witten, ``Anti de Sitter Space And Holography,'' hep-th/9802150,
Adv. Theor. Math. Phys. {\bf 2} (1998) 253.}
In this particular example, because the boundary of $\AdS_5$ has
positive scalar curvature $R$, one expects the Coulomb branch to be suppressed
because of an $R\phi^2$ interaction.  We will see this behavior below as 
a divergence in the action or energy
when the large threebrane approaches the boundary.
We will also generalize the discussion to consider instead of $\AdS_5$
a more general negatively curved Einstein five-manifold, whose boundary
may have negative $R$.  In that case, the threebrane action or energy
will go to $-\infty$
when the threebrane approaches the boundary, reproducing the expected
unstable behavior of the conformal field theory on a manifold of negative $R$.

Though we have framed this discussion for $\AdS_5$, we can similarly
probe the approach to the Coulomb branch in any of the $\AdS_{D+1}$ examples
by considering the behavior of a  large $D-1$-brane.  Many of the properties
are independent of $D$ for $D>2$, but special things happen for $D=2$,
where the $D-1$-brane is a string.  As we will see, the effective
theory of the large string
is a Liouville theory.  By studying it, we will be able to understand
the singularity of the Higgs branch for the $\AdS_3$ examples.

\subsec{Analysis Of The Large Brane}

With the motivation that we have just explained, and following
 Maldacena, Michelson and Strominger \mms, we
study the dynamics of a large $D-1$ brane in $\AdS_{D+1}$.  
We consider a large brane carrying charge $q$
under the background antisymmetric tensor field.  The brane is
really moving on $AdS_{D+1}\times {\bf Y}$ for some ${\bf Y}$, but
for the moment we can ignore the motion on ${\bf Y}$.  

We let $A$ denote the volume of $\partial W$, and we let $V$ denote
the volume of $W$.  The brane action has two terms, one a positive
multiple of $A$ coming from the brane tension, and the other a
negative multiple of $V$ coming from the ``Wess-Zumino coupling'' of
the brane to the background antisymmetric tensor field.  In flat
spacetime, for a sufficiently large brane one has $V\gg A$, so the
action of a sufficiently large brane in the presence of a constant
antisymmetric tensor field strength is negative.  Hence a sufficiently
large brane grows indefinitely.  A constant antisymmetric tensor field
in flat spacetime would relax to zero by nucleation of branes; for
example, this mechanism leads to the periodicity of the $\theta$ angle
in two dimensional QED.  In $\AdS$ space, this energetics is more
delicate because $V$ is proportional to $A$.  The BPS case is
precisely the case that the leading volume and surface terms cancel.
Much physics is contained in the subleading terms that do not cancel.

We start by writing down the metric of $\AdS$ space in the following form:
\eqn\metricads{ds^2=r_0^2(dr^2+\sinh^2 r d\Omega^2).}
Here $d\Omega^2$ is the round metric on $\bS^D$.  The topology of the
$D$ dimensional worldvolume of the brane is $\bS^D$ and it is located at
$r(\Omega)$.

As a preliminary for finding the effective action of the brane we
calculate $V$ and $A$.  We easily find that the volume enclosed by the brane is
\eqn\volume{\eqalign{
V=& r_0^{D+1}\int d^D\Omega \int_0^r dr' \sinh^D r' ={ r_0^{D+1} \over
2^D} \int d^D\Omega \int_0^r dr' \left(e^{Dr'}-De^{(D-2)r'} +
\CO(e^{(D-4)r'}) \right)\cr 
&={\cases{
{r_0^{D+1} \over 2^D} \int d^D\Omega \left( {1 \over D} e^{Dr}-{D
\over D-2} e^{ (D-2)r} + \CO(e^{(D-4)r}) \right) & for $D>2$; \cr   
{r_0^3 \over 4 }\int d^2\Omega \left( {1 \over 2} e^{2r}-2r +
\CO(e^{-2r})\right) &  for $D=2$.\cr }}}}
Here $d^D\Omega$ is the volume element of a unit sphere.

Similarly, the surface volume of the brane is
\eqn\area{\eqalign{
A=&r_0^D \int d^D \Omega {1 \over \sqrt g }\sqrt{ \det_{\alpha\beta}
\left(\sinh^2r g_{\alpha\beta} + \partial_\alpha r \partial_\beta r
\right)} \cr = &
{r_0^D \over 2^D} \int d^D \Omega \left( e^{Dr} - De^{(D-2)r} + 2
e^{(D-2)r} (\partial r)^2 +\CO(e^{(D-4)r}) \right) \cr}}
where $g_{\alpha\beta}$ is the round metric on the unit sphere.

In the above formulas, we can think of ``$r$'' as an effective field
on the large brane.  Before combining these formulas to compute the action of 
a large brane as a function of this field, 
we will put them in a more covariant form.  This will make
it clear how $r$ transforms under Weyl transformations in the boundary
theory, and why.  

The boundary of $\AdS_{D+1}$ has a natural conformal structure but not
a natural metric.  Let $ds^2=g_{ij}dx^idx^j$ be an arbitrary metric on
the boundary in its conformal class.  Here the $x^i$, $i=1,\dots,D$
are an arbitrary set of local coordinates on the boundary.  
There is then (see Lemma 5.2 in \ref\graham{C. R. Graham and R. Lee, 
``Einstein Metrics With Prescribed Conformal Infinity On The Ball,''
Adv. Math. {\bf 87} (1991) 186.}) a unique way to extend the $x^i$
to  coordinates on $\AdS_{D+1}$ near the boundary, adding an
additional coordinate  $t$
that vanishes on the boundary,  such that the metric in a neighborhood
of the boundary is 
\eqn\kilop{ds^2={r_0^2\over t^2}\left(dt^2+\widehat g_{ij}(x,t)dx^idx^j\right)
      }
 with \eqn\kkilop{\widehat g_{ij}(x,0)=g_{ij}(x).}
Moreover, one can use the Einstein equations to determine the behavior of
$\widehat g_{ij}$ near $t=0$.  One has
\eqn\milop{\widehat g_{ij}(x,t)=g_{ij}(x) -t^2P_{ij}+{\rm higher~orders~in}~t,}
where for $D>2$
\eqn\rullo{P_{ij}={2(D-1)R_{ij}-g_{ij}R\over 2(D-1)(D-2)},}
which implies
\eqn\ikolo{g^{ij}P_{ij}= {R\over 2(D-1)}.}
This last formula
 is the only property of $P$ that we will need.  For $D=2$, \rullo\
is no longer valid, but \ikolo\ is.  (For $D=2$, the Einstein equations
do not determine the trace-free part of $P$ in terms of local data near
the boundary.)

In this formulation, we can see how $t$ transforms under Weyl rescalings
of the boundary metric.  If we take $g_{ij}\to e^{2\omega}g_{ij}$,
then $t$ must be transformed to maintain the properties \kilop, \kkilop.
Clearly, this requires
\eqn\uncu{t\to e^\omega t+\dots,}
where the ellipses are terms of higher order near $t=0$.  This means
that $t$ (if corrected by adding higher order terms to remove
the ellipses) 
is a field of conformal dimension $-1$.  The canonical dimension
of a scalar field is $(D-2)/2$ for $D>2$, so we should expect that
a scalar field $\phi$ 
of canonical dimension will be $\phi\sim t^{-(D-2)/2}$ for $D>2$.
Instead, in classical field theory in $D=2$, an ordinary
 scalar field is dimensionless.  To make a field $t$ of dimension $-1$
as a function of a two-dimensional real scalar field $\phi$, $\phi$ must
be a Liouville field (with a coupling $\phi R$ to the worldsheet
curvature $R$) and $t$ must be written as a real exponential of $\phi$.

Before computing $V$ and $A$ in the covariant approach, we set $t=2e^{-r}$,
so that the metric becomes
\eqn\plxo{ds^2={r_0^2}\left(dr^2+{e^{2r}\over 4}g_{ij}dx^idx^j-P_{ij}dx^idx^j
+\CO(e^{-2r})\right).}
This will make the comparison with the formulas \volume\ and \area\ more
transparent.  Also, as suggested in the last paragraph, we can write
$r$ in terms of a canonical scalar field $\phi$ for $D>2$, or a Liouville
field $\phi$ for $D=2$, by
\eqn\redefr{r=\cases{
{2 \over D-2} \log \phi + {1\over (D-1)(D-2)} \phi^{-{4 \over D-2} }R
& for $D>2$ \cr
\phi +e^{-2\phi}\phi R & for $D=2$.}}
The leading terms in this formula, which correspond to $t={\rm const}\,
\phi^{-2/(D-2)}$ for $D>2$, and $t={\rm const}\,
e^{-\phi}$ for $D=2$, were explained
in the last paragraph.  The correction terms in \redefr\ can presumably
be calculated by computing the higher order terms in \uncu\ and then
seeing how $t$ can be expressed in terms of a canonical scalar field
for $D>2$, or a Liouville field for $D=2$.  We will not be as systematic
as this because the correction terms are unimportant for BPS branes,
which are our main interest.  We have merely determined the coefficients
of the correction terms to make the formulas below conformally invariant
even for the non-BPS case.

Using \plxo, we can compute that the volume enclosed by the brane, 
computed with an arbitrary metric
on the boundary in its conformal class and using the associated $r$-function,
is
\eqn\volumeg{
V=\cases{
{r_0^{D+1}  \over 2^D} \int d^Dx\sqrt g \left( {1 \over D} e^{rD}-{1\over
(D-1)(D-2)} e^{ (D-2)r} R + \CO(e^{(D-4)r}) \right) & for $D>2$; \cr
{r_0^3 \over 4 } \int d^Dx\sqrt g \left( {1 \over 2} e^{2r}- r R +
\CO(e^{-2r})\right) &  for $D=2$ \cr }}
This reduces to \volume\ when the boundary is a unit sphere with a round
metric, for which $R=D(D-1)$.  
In terms of the canonical scalar or Liouville field $\phi$,
we have
\eqn\volumegr{V=\cases{
{r_0^{D+1}  \over D2^D} \int d^Dx\sqrt g \left(\phi^{2D\over D-2} +
\CO(\phi^{2(D-4)\over D-2})\right) & for $D>2$; \cr   
{r_0^3 \over 8 } \int d^Dx\sqrt g\left( e^{2\phi}+ \CO(e^{-2\phi })\right) &
for $D=2$. \cr }}  
Likewise, in the same generality, the
surface volume of the boundary is 
\eqn\areag{\eqalign{
A= &{r_0^D \over 2^D} \int d^Dx\sqrt g \left( e^{Dr} - {1\over D-1}
e^{(D-2)r}R + 2e^{(D-2)r} (\partial r)^2 +\CO(e^{(D-4)r}) \right) .\cr}}
In terms of $\phi$, this is
\eqn\intermso{A
=\cases{
{r_0^D \over 2^D} \int d^Dx\sqrt g \left( \phi^{2D\over D-2}+ {8 \over
(D-2)^2} [(\partial\phi)^2 + {D-2\over 4(D-1)}\phi^2 R]
+\CO(\phi^{2(D-8)\over D-2})\right) & for $D>2$; \cr
{r_0^2 \over 4} \int d^Dx\sqrt g \left( e^{2\phi} + 2[ (\partial\phi)^2
+\phi R] - R  +\CO(e^{-2\phi})\right) & for $D=2$. }}

One of the advantages of the covariant derivation that we have given
is that these formulas are not restricted to branes near the boundary
of $\AdS_{D+1}$.  One can replace $\AdS_{D+1}$ with an arbitrary
Einstein manifold $W$  of negative curvature and conformal boundary $M$.
The same formulas hold, with the same derivation, for
the area and Wess-Zumino coupling of a large brane that is near $M$,
or, if $M$ is not connected, near any component of $M$.

Now, let us combine these formulas and study the behavior of branes.
The action of a  brane that couples to the background antisymmetric
tensor field with charge $q$ receives two contributions.  One
term, arising from the tension of the brane $T$, is $TA$.
The interaction of its charge with the background gauge fields leads
to a term proportional to $V$.  The whole action is
\eqn\eqnprop{\eqalign{S=&T(A-{qD\over r_0} V) = \cr
=&\cases{
{Tr_0^D \over 2^D} \int \sqrt g\left((1-q)\phi^{2D\over D-2} +
{8\over (D-2)^2}[(\partial\phi)^2 + {D-2\over 4(D-1)}\phi^2 R]
+ \CO(\phi^{2(D-4) \over D-2}) \right) & for $D>2$;\cr 
{Tr_0^2\over 4}\int \sqrt g \left((1-q)e^{2\phi} +2 [(\partial\phi)^2
+\phi R] -R  + \CO(e^{-2\phi }) \right)& for $D=2$ .\cr}}}

The brane approaches the boundary in the limit $\phi\to\infty$.
For large $\phi$, the dominant term is the first one, $(1-q)\phi^{2D/(D-2)}$
for $D>2$
or $(1-q)e^{2\phi}$ for $D=2$.  
For $q>1$, the action goes to $-\infty$ for $\phi\to
\infty$ and the system is unstable against the emission of branes.
However, in a supersymmetric $\AdS$ vacuum, there will never
be a brane with $q>1$ as this violates a BPS bound.  In a supersymmetric
theory, we will have either BPS branes of $q=1$ or non-BPS branes of $q<1$.

If $q<1$,  the leading order term in \eqnprop\ is a
potential which tries to contract the brane.  In this case, the effect
of the tension of the brane is larger than the force due to the
charge.  The brane tends to contract and, if we are
in $\AdS_{D+1}$, eventually annihilates. If
$\AdS_{D+1}$ is replaced by a more general Einstein manifold $X$, the
non-BPS brane might conceivably contract to a stable minimum of the action.

For BPS branes, $q=1$ and the effective action becomes
\eqn\eqnpropbps{S =\cases{
{Tr_0^D \over 2^{D-3}(D-2)^2 } \int \sqrt g\left(
(\partial\phi)^2 + {D-2\over 4(D-1)}\phi^2 R
+ \CO(\phi^{2(D-4) \over D-2}) \right) & for $D>2$;\cr 
{Tr_0^2\over 2}\int \sqrt g \left((\partial\phi)^2 +\phi
R -\half R  + \CO(e^{-2\phi }) \right)& for $D=2$.\cr}}
The BPS term has caused the ``potential energy'' term to cancel
out, leaving an action that is for $D>2$ the minimal conformally
invariant kinetic energy of a scalar field, and for $D=2$ is the minimal
conformally invariant kinetic energy of a Liouville field.

Do BPS $D$-branes exist?  For the usual $\AdS$ models
with $D>2$, they generally do.  For example, there are BPS three-branes
in the case of $\AdS_5\times \S^5$.  The reason is that the usual $D>2$
models are constructed from the near horizon geometry of a system
of parallel branes of just one type.  In such a model, 
a probe brane of the type used in building the vacuum is
BPS.  On the other hand, the $\AdS_3$ or $D=2$ examples are constructed
{}from more than one type of brane, and then, as we have described in
section 2, for generic values of the moduli there are no BPS probe
branes.

Now, let us focus on what happens when BPS branes do exist.  In string
theory or $M$-theory on $\AdS_{D+1}$ (or more exactly
$\AdS_{D+1}\times Y$ for some $Y$) the boundary is $\S^D$, and its
conformal structure is that of the round metric on a unit sphere.  If
$S$ is evaluated with the round metric, which has $R>0$, it is
manifestly positive definite and divergent for $\phi\to\infty$.  Since
$S$ is conformally invariant, this is true for any metric in the same
conformal class.  Thus, on $\AdS_{D+1}$, there is no instability from
emission of a large brane.  This is in accord with \mms, where it was
shown that there is no such instability except for $D=1$ (a case that
we are not treating in the present paper).

On the other hand, suppose that we replace $\AdS_{D+1}\times Y$ by 
$W\times Y$, where $W$ is
a more general
Einstein manifold $W$ of conformal boundary $M$.  The conformal class of
metrics on $M$ may admit a representative with negative $R$.  If so, $S$ is
not positive definite and 
the string theory on $W\times Y$ is unstable against emission of a large
$D$-brane that approaches the boundary.  

These results agree with   field theory expectations in the important
case that $D=4$ and $Y$ is, for example, $\S^5$.  Then the boundary
theory is $SU(N)$ super Yang-Mills theory (or $U(N)$ if we include the 
singleton field
on the boundary).  If we formulate this theory on a four-manifold $M$ in
a conformally invariant fashion,
and try to Higgs the $SU(N)$ gauge group to $SU(N-1)\times U(1)$ by
giving an expectation value to a component $\phi$ of the scalar
fields of the theory, then the conformally invariant kinetic energy
for $\phi$ is precisely the functional $S$ obtained above.  Hence,
at the field theory level, we expect an instability when $M$ has negative
scalar curvature (and more generally when the conformally invariant
functional $S$ is not positive semi-definite).  In this way, string
theory reproduces the instability that is evident in the field theory.

At this level, the results do not yet seem to distinguish $D=2$ from
$D>2$.  The conformal field theory of one of the $D=2$ examples,
just like those with $D>2$, is stable if formulated on a boundary
with $R>0$, but not if $R<0$.  However, $D=2$ is clearly a delicate
case, since the growth of the action as a BPS brane approaches
the boundary is much slower -- linear in $r$ for $D=2$ and exponential
for $D>2$.  This suggests that we should look at $D=2$ more closely.

To understand what is special about $D=2$, it helps to consider a
Hamiltonian formalism and to consider the energy of a large brane
rather than its action, as considered up to this point.  Thus, we are
now considering the large brane as a physical object, whereas so far,
our branes (as Euclidean space objects) were really instantons.  For
this, we go to the Lorentz signature version of $\AdS_{D+1}$, so that
the boundary is now $\R\times\S^{D-1}$ (with $\R$ parametrizing the
time direction) rather than $\S^D$.  The formula for the large brane
action $S$ is still valid, and from it we can read off an effective
Hamiltonian for the large brane.  In particular, the energy has an
$R\phi^2$ or $R\phi$ term for $D>2$ or $D=2$.  The main point is that
for $D>2$, $\R\times \S^{D-1}$ has $R>0$, and hence in the Hamiltonian
formulation, the energy diverges as a brane approaches the boundary.
But for $D=2$, the boundary is $\R\times \S^1$ and has zero scalar
curvature.  Hence, the energy of the brane does not grow as the brane
is stretched.

We can make a more precise statement using the fact that given the
form of $S$, the effective theory on the brane is a Liouville theory
(for a review see
\ref\liouvillen{N.~Seiberg, ``Notes On Quantum Liouville Theory And
Quantum Gravity,'' Prog. Theor. Phys. Suppl. {\bf 102} (1990) 319.}).
The normalizable states of Liouville
the theory form a continuum above a certain
positive threshold.  In order to find the properties of the Liouville
theory of the long string \eqnpropbps, it is convenient to rescale
$\phi$ such that its kinetic term is canonical.  Then, the coupling to
the two dimensional curvature $R$ leads to an improvement term in the
stress tensor with coefficient
\eqn\qsemva{Q=\sqrt{4\pi Tr_0^2}}
leading to a threshold at
\eqn\emin{E_0={Q^2 \over 4} = \pi Tr_0^2 .}
We conclude that a BPS brane in $\AdS_3$ can be stretched to infinity
with a finite cost in energy \mms.  

This finite non-zero energy is in
accord with the fact that in the Euclidean version, a large brane
approaching the boundary of $\AdS_3$ has an action that diverges for 
$r\to\infty$,
albeit slowly.  Such a large brane is an ``instanton.''  A finite action
instanton describes a tunneling process to a state of degenerate
energy.  Since we are trying to get to a state of finite energy -- with
a large brane in real time -- there
cannot be a finite action instanton.

As promised in the introduction, we have found a dangerous region in
the spacetime conformal field theory of the D1/D5 system where it is
described by an effective Liouville theory.  
We claim that, if the long string is a D1 brane, then
 the behavior for large $\phi$ corresponds to
the small instanton singularity.  
The evidence for this claim is first that, as we explained at the outset,
intuitively emission of a long string is a way to reduce the charges of
the system.  Second,
the dangerous large $\phi$ region occurs precisely when the small instanton
is BPS; when $\theta$ angles are turned on,
as reviewed in section 2.2, to suppress the small instanton singularity,
then the long string has $q<1$ and the large $\phi$ region is suppressed
by an exponential potential.
Finally, we note the following important check of the identification
of the small instanton singularity with the large $\phi$ region.  The small 
instanton
singularity of the D1/D5 system depends only on $Q_5$ and not $Q_1$
(since the singularity when an instanton is small is entirely independent
of how many non-small instantons there are).  This is entirely in accord
with the fact that, if the long string is a D1 string, then the Liouville
action $S(\phi)$ depends on $Q_5$ and not $Q_1$.

To be more precise about this, 
to describe the small instanton singularity of the D1/D5 system
of charges $(Q_1,Q_5)$, we use a long $D$-string plus
a D1/D5 system of charges $(Q_1-1,Q_5)$.  
The  action $S$ actually describes the motion of the long $D$-string on
$\AdS_3$, and 
must be supplemented by additional terms describing its motion
 in other dimensions of $\AdS_3\times \S^3\times \bX$.  In particular,
the position of the long string on $\bX$ corresponds, in the other description,
to the position of the small instanton on $\bX$.

Like the D1/D5 system, the 
 other singularities described in the instroduction
have an analogous tubelike description using
other long BPS strings.
Whenever there is a BPS string, the D1/D5 system (and its $U$-dual cousins)
has a continuous spectrum at energies
above \emin, though the spectrum is discrete at energies smaller
than this.   The states in
the continuum are states that contain a long string.  States below threshold
have wavefunctions that vanish exponentially in the part of phase space
where there is a long string.
 
Likewise, the operators in the boundary conformal field
theory have a discrete spectrum of
conformal dimensions up to 
\eqn\deltamin{\Delta_0={E_0\over 2} =\half\pi Tr_0^2.}
The operators creating the long string and the fluctuations on it have
continuous dimensions starting at $\Delta_0$.

Conformal field theories with a continuous spectrum of dimensions are
most familiar in cases when the target space is non-compact.  The
example of a single non-compact boson is well known.  In this case the
continuum is associated with the arbitrary momentum of the boson.
Momentum conservation guarantees that in the operator product
expansion of two operators there is only one value of the momentum,
and therefore there is a discrete sum over the operators which appear
there.  When the target space is non-compact and is not
translationally invariant, there is a continuum of operators in the
operator product expansion.  We now see that the D1/D5 system, which
appears to have a compact target space $\CM$ (the moduli space of
$U(Q_5)$ instantons with instanton number $Q_1$), also has a
continuous spectrum, albeit above a gap.  This is possible because
$\CM$ has singularities.  The description of the conformal field
theory as a sigma model is not very useful near singularities of
$\CM$.  The good description of the behavior near the singularities is
in terms of a long string plus a residual system of lower charges.

For the D1/D5 system, the long string is a D-object; it can be any
collection of D1, D3 and D5 branes which forms a BPS string.
Therefore, from the point of view of a conformal field theory
description of the D1/D5 system (such as the one in
\ref\berk{N. Berkovits, C. Vafa, and E. Witten, ``Conformal Field
Theory Of AdS Background With Ramond-Ramond Flux,'' hep-th/9902098.}),
the long string effect is non-perturbative.  The same system, in a different
region of its parameter space, has a more natural description as an
NS1/NS5 system, built from NS5 branes and weakly coupled fundamental
strings.  In this region, the lightest string which can escape to
infinity is a fundamental string.  Hence our instanton is in this case
a genus zero worldsheet instanton.  It contributes at string tree
level, but its contribution is non-perturbative in $\alpha'= {1\over
2\pi T} $.

\bigskip\noindent{\it The String Coupling Constant}

The boundary conformal field theory is a conformal field theory in
which the string coupling constant $\lambda_{eff}(\phi)$ blows up as
$\phi\to\infty$, that is, as the long string goes to infinity.  This
is because of the $\phi R$ coupling.  The divergence of $\lambda$ is
actually the reason that the action for the instantonic long string --
wrapped around the boundary of Euclidean $\AdS_3$ -- diverges as
$\phi$ goes to infinity.  (The instanton amplitude is proportional to
$\exp(-S)\sim \exp(-\sqrt{4\pi Tr_0^2}\phi)$, which we can interpret
as $1/\lambda_{eff}^2(\phi)$.)  The divergence of
$\lambda_{eff}(\phi)$ also causes the partition function of the
boundary conformal field theory to be divergent for $\phi\to \infty$
if formulated on a Riemann surface of genus $>1$.  We have already
noted above this consequence of the $R\phi$ coupling for the case that
$R<0$.

If we formulate the boundary CFT in genus zero (for instance on the
boundary of $\AdS_3$), then because $R>0$, the partition function
converges.  However, one would suspect that the natural operators of
the boundary CFT will have an exponential dependence on $\phi$, and
that some correlation functions will diverge in integrating over
$\phi$.  Thus, some correlation functions of the boundary CFT of the
D1/D5 system are likely to receive divergent contributions from the
small instanton singularity, as well as singularities from partial
un-Higgsing.  Of course, as reviewed in section 2, these divergent
contributions can be avoided if one turns on suitable theta angles and
RR fields so that there are no BPS-saturated strings.  In any event,
determining the $\phi$ dependence of operators and showing that there
really are divergent contributions to genus zero correlation functions
is beyond the scope of the present paper.

For some applications of conformal field theories of this kind,
the divergence of the effective string coupling for $\phi\to\infty$
is very important.
For example, if we set $Q_5=2$, then the small instanton singularity
is in codimension four and looks like $\R^4/\Z_2$.  Specializing to this
case, the analysis shows that in terms of the right variable (which
is the long string position) the effective string coupling in
$\R^4/\Z_2$ conformal field theory (with zero theta angle) goes
to infinity as one approaches the singularity.  This makes it possible
for Type IIA string theory on $\R^4/\Z_2$ to have nonperturbative behavior
(enhanced gauge symmetry)
no matter how small the bare string coupling constant may be.

\bigskip\noindent{\it Change In Central Charge}

Because of the $\phi R$ term in the action \eqnprop, the Liouville
theory of the long string has central charge
\eqn\csemicl{c=3Q^2= 12\pi Tr_0^2.}
We can interpret the long string as carrying away that amount of $c$
out of the full conformal field theory of the D1/D5 system.

For the D1/D5 system,
using $Tr_0^2={Q_5\over 2\pi} $, the threshold \deltamin\ and the central
charge \csemicl\ become
\eqn\delmc{\eqalign{ \Delta_0&= {Q_5 \over 4} \cr
c&=6Q_5.}}
The boundary conformal field theory of the
full system is known to have $c_{full}=6Q_1Q_5$.  
When a string is emitted, the remaining system has charges $(Q_1-1,Q_5)$,
so its central charge is reduced by $6Q_5$.  
This coincides with the central charge of
the long string itself, so we see that the string whose emission
reduces $Q_1$ by one carries with it the remaining value of $c$.  This
is compatible with the claim that in a certain region of its phase
space, the $(Q_1,Q_5)$ system behaves like a long string plus
a $(Q_1-1,Q_5)$ system.

The formulas in \delmc\ are, however, only semiclassical approximations.
A more precise derivation will be presented in section 4.

\newsec{Exact Analysis Of The Long String}

\def\CM{{\cal M}}

The preceding analysis of the central charge and the gap is valid only
for  $Q_1, Q_5 \gg 1$, where the the long string does not
affect the background geometry significantly.  We now present a more
complete analysis, for the weakly coupled NS5/NS1 system.  It is valid
for $Q_1\gg 1$ but without assuming that $Q_5$ is large.

We will explain several approaches.  To start with, we use the RNS
formalism as in \refs{\gks-\ks}.  The system has short strings
(ordinary perturbative excitations) and the long strings that we have
been considering.  Using a coordinate system with the metric
\eqn\adsm{ds^2= dx^2+ e^{2x}d\gamma d\bar \gamma, }
the worldsheet Lagrangian is
\eqn\wsl{\int\left(\partial x\bar\partial x + e^{2x}\bar \partial \gamma
\partial \bar \gamma\right)d^2z.}
By introducing auxiliary fields $\beta$, $\bar \beta$, one can
write an equivalent Lagrangian
\eqn\nsl{\int \left(\partial x\bar\partial x+\beta\bar\partial\gamma
+\bar\beta\partial\bar\gamma -e^{-2x}\bar\beta\beta\right) d^2z.}
A long string can be described by the solution of the worldsheet
equations of motion 
\eqn\longs{\eqalign{
&x=x_0\gg 1 \cr 
&\gamma(z,\bar z)=z \cr 
&\bar \gamma(z,\bar z)=\bar z.}}  
Since we take $x_0\gg 1$, we can treat the $e^{-2x}\bar\beta\beta$
term in \nsl\ as a small perturbation, and use the free field
representation of $SL(2)$ conformal field theory as in \gks.
Furthermore, the calculation of the central charge of the space-time
Virasoro algebra in \gks\ applies directly to the long string rather
than to the whole system.  In particular, the value of the central
charge found there, namely
\eqn\centralgks{c=6Q_5\oint dz {\partial \gamma\over \gamma} = 6Q_5,}
coincides with our semiclassical result \delmc.  More generally, if
$\gamma$ in \longs\ is replaced by $\gamma=z^m$, which corresponds to
$m$ coincident long strings, \centralgks\ becomes \gks\
\eqn\qgeneral{c=6Q_5m.}
In the rest of this paper we will be mostly concerned with a single
long string; i.e.\ $m=1$.

This calculation is
to be contrasted with the calculation of the central charge of the
full theory including the short strings \refs{\dbort,\ks}, which
receives contributions from disconnected diagrams.  For more details
about these two calculations and the relation between them see \ks.

\def\NN{{\cal N}}
We do not, however, know how to compute the value of the threshold
{}from this point of view.  We will therefore present two additional
calculations which determine both the central charge of the long
string and the threshold.  The first computation is based on the
covariant RNS description with ghosts, and the second on a physical
gauge.  (The first of the two calculations is somewhat intuitive
and needs to be put on a firmer footing than we will achieve.)

We start by considering the bosonic string on $\AdS_3\times {\bf M}$
($\bf M$ might be, for instance $\S^3\times \T^{20}$).
In the covariant formalism the worldsheet stress tensor is
\eqn\stresswsb{T_{worldsheet} = T_{SL(2)}+T_{\bf M} + T_{ghosts}}
where $T_{SL(2)}$ is constructed out of a bosonic $SL(2)_{k}$ WZW
theory.  $T_{\bf M}$ is the stress tensor of the conformal field
theory on ${\bf M}$ and $T_{ghosts}$ is constructed out of the $b,c$
ghosts.  The central charges of these stress tensors are
\eqn\centchargesb{\eqalign{
&c_{SL(2)} = {3k \over k-2} \cr
&c_{\bf M}= 26 - {3k \over k-2}  \cr
&c_{ghosts}= -26.}}
Note that the total central charge $c_{worldsheet} = c_{SL(2)}+ c_{\bf
M} +c_{ghosts}$ vanishes.

In the Wakimoto representation, using the fields in \nsl,
 the three $SL(2)$ currents of the
bosonic level $k$ WZW theory are
\eqn\curr{\eqalign{
&J^-= \beta \cr
&J^3= \beta\gamma + \half \sqrt{2(k-2)}\partial x \cr
&J^+= \beta\gamma^2 + \sqrt{2(k-2)}\gamma \partial x + k\partial
\gamma.}} 
Evaluating them on the classical solution of the long string we find
\eqn\currc{\eqalign{
&J^-= 0\cr
&J^3= 0\cr
&J^+= k.}}
Therefore, it is natural to impose a lightcone-like gauge\foot{Related
ideas were explored in
\ref\hamred{E. Verlinde, H. Verlinde and H. Ooguri, unpublished;
J. Maldacena, unpublished; M.~Yu and B.~Zhang, ``Light cone gauge
quantization of string theories on AdS(3) space," hep-th/9812216.}.} 
$J^+= k$.  More precisely, we study the BRST cohomology of our
string after expanding the fields around this classical configuration.
To compute this cohomology, we imitate the proof of the no ghost theorem
for the bosonic string in flat space, as presented
in \ref\polchinski{J. Polchinski, {\it String Theory} (Cambridge University
Press, 1998), Vol. 1, section 4.4.}.  Because $\gamma$ only appears
in the Lagrangian 
via a term proportional to $\bar\partial\gamma$, there is a symmetry
$\gamma\to \gamma+z$.  We define an operator $N$ (analogous to
$N^{\rm lc}$, introduced in \polchinski, eqn. (4.4.8)) that acts as
$\oint J^3$ on all modes of $\beta,$ $\gamma$, and $\phi$, except the
mode $\gamma\sim z$ on which it acts trivially.  Because of that one mode,
the operator $N$ does {\it not} commute with
$Q_{BRST}$.  The term in $Q_{BRST}$ with the most negative $N$ charge
is $\tilde Q_{BRST}=
\oint c J^- $.  The equation $Q_{BRST}^2=0$ reduces, for the term
of most negative $N$ charge, to $\tilde Q_{BRST}^2=0$.  The same
argument as in the usual proof of the no ghost theorem in flat space
shows that the cohomology of $Q_{BRST}$ is the same as that of $\tilde
Q_{BRST}$.  Also, as in flat space, taking the $\tilde Q_{BRST}$
cohomology effectively removes from the Hilbert space the fields
$\beta$ and $\gamma$ along with the
conformal ghosts $b$ and $c$.  The cohomology consists of all states
built by acting on the ground state with the other fields.\foot{A very
similar no ghost 
argument for the $SL(2,\R)$ WZW model, using a free field representation,
can be found in \ref\bars{I. Bars, ``Ghost-Free Spectrum Of A Quantum
String In $SL(2,\R)$ Curved Space-Time,'' Phys. Rev. {\bf D53} (1996) 3308,
hep-th/9503205; ``Solution Of The $SL(2,\R)$ String In Curved Spacetime,''
in {\it Future Perspectives In String Theory} (Los Angeles, 1995), 
hep-th/9511187.}.  For another approach to this model and its no ghost
theorem, see \ref\perry{J. Balog, L. O'Raifeartaigh, P. Forgacs, and A. Wipf,
``Consistency Of String Propagation On Curved Space-Times: An $SU(1,1)$
Based Counterexample,'' Nucl. Phys. {\bf B325} (1989) 225; P. M. S.
Petropoulos, ``Comments On $SU(1,1)$ String Theory,'' Phys. Lett. {\bf B236}
(1990) 151; S. Hwang, ``No-Ghost Theorem For $SU(1,1)$ String Theories,''
Nucl. Phys. {\bf B354} (1991) 100;  J. M. Evans, M. R. Gaberdiel, and
M. J. Perry, ``The No-Ghost Theorem And Strings On $\AdS_3$,''
hep-th/9812252.}.}
 
The procedure just sketched is similar to that of the Hamiltonian
reduction of $SL(2)$ as in \ref\bershoo{M. Bershadsky and H. Ooguri,
``Hidden $SL(N)$ Symmetry In Conformal Field Theories,''
Commun. Math. Phys.  {\bf 126} (1989) 49.}, but the problem we are
studying and therefore also the details of the construction are
somewhat different.  We are computing the physical state spectrum for
a bosonic string on $\AdS_3\times {\bf M}$, so in particular we have
the bosonic string $b,c$ ghosts of spins $-2$ and $1$ together with
additional variables, while in Hamiltonian reduction a different
problem is solved, so the ${\bf M}$ degrees of freedom are absent, and
different ghosts are used.  Our gauge fixing condition is therefore
also somewhat different, though the computation of the cohomology ends
up being similar.

Now we want to find the spacetime stress tensor.  It acts on the
BRST cohomology, which we have identified as the cohomology of the
operator $\oint cJ^-$.  In this representation, the spacetime stress tensor
must be an operator that commutes with $\oint cJ^-$.
For this to be so, $J^-$ must have conformal dimension 2 relative
to the spacetime stress tensor (though its dimension with respect to the
worldsheet stress tensor is 1).
Also, the current $J^+$, which is expanded around a constant, should have
conformal dimension zero.  These
dimensions are obtained by twisting the stress tensor \stresswsb\ to 
\eqn\stresstb{T_{total}= T_{worldsheet} + \partial J^3 .}
We interpret $T_{total}$ as the space-time stress tensor of the theory
on the long string.  Computing the central charge of \stresstb\
using $J^3(z)J^3(w) \sim - {k/2 \over (z-w)^2}$ we find
\eqn\ctotalb{c_{total}= 6k.}

The physical degrees of freedom living on the long string include the
modes of the sigma model on ${\bf M}$ describing the location of the
string in ${\bf M}$, and the $x$ boson which we now identify as the
$\phi$ field of section 3.  The central charge \ctotalb\
is obtained because the boson $\phi$ becomes a Liouville field with an
improvement term $Q$.  In order to find the improvement term we
equate the two expressions for the central charge
\eqn\centequ{ 26-{3k\over k-2}+1+3Q^2=6k,}
and find
\eqn\qvalue{Q=(k-3)\sqrt{2\over k-2}.}
Because of this improvement term the system has a gap
\eqn\exactgapb{\Delta_0= {Q^2\over 8} = {(k-3)^2 \over 4(k-2)}.}

The two computations we have so far described give the same formula
for the central charge in the spacetime conformal field theory,
but give seemingly very different formulas for the spacetime stress tensor.
How are they related?  This question can be partly answered as follows.
In the classical approximation of large $k$ we ignore the ghosts and set the 
worldsheet stress tensor \stresswsb\ to zero
\eqn\stresswsbn{0 = {-(J^3)^2 + J^+J^- \over k} +T_{\bf M}}
(we neglect the shift of $k$ by 2).  We substitute $J^+=k$ and solve
for $J^-$ 
\eqn\solvjm{J^-={(J^3)^2 \over k}   - T_{\bf M}.}
Equation (2.37) in \gks\ gives the spacetime Virasoro generators
\eqn\virg{L_n=\oint dz \left[nJ^-\gamma^{n+1} - (n+1)J^3\gamma^n
\right] = \oint dz \left[ - J^-\gamma^{n+1} - \half (n+1) \sqrt{2k}
\partial x \gamma^n \right] ,}
where we used the large $k$ limit of \curr.  Substituting
\solvjm\ in \virg\ and ignoring terms which vanish for large $k$ we
find 
\eqn\virgs{L_n= \oint dz \left[\left(T_{\bf M} - \half (\partial
x)^2\right)\gamma^{n+1} - \half (n+1) \sqrt{2k} \partial x \gamma^n
\right] .} 
Using our gauge choice $\gamma=z$
\eqn\virgsg{L_n= \oint dz \left[T_{\bf M} - \half (\partial
x)^2 +\half \sqrt{2k} \partial^2 x\right] z^{n+1}.}
Therefore, the spacetime stress tensor is
\eqn\spacetg{T_{spacetime}= T_{\bf M} - \half (\partial x)^2 +\half
\sqrt{2k} \partial^2 x,} 
whose central charge for large $k$ is $c=6k$ in agreement with the
exact answers.  Note also that the boson $x$ acquires an improvement
term as above.

We now extend this analysis to superstrings on $\AdS_3\times {\bf
S}^3\times {\bf X}$ in the RNS formalism.  The worldsheet stress
tensor is
\eqn\stressws{T_{worldsheet} = T_{SL(2)}+T_{SU(2)} + T_{\bf
X}+T_{ghosts}.} 
$T_{SL(2)}$ is constructed out of a bosonic $SL(2)_{Q_5+2}$ WZW theory
and three free fermions $ \psi^{\pm}$ and $\psi^3$.  $T_{SU(2)} $ is
constructed out of a bosonic $SU(2)_{Q_5-2}$ WZW and three free
fermions, $\chi^a$.  $T_{\bf X}$ is the stress tensor of the
superconformal field theory on ${\bf X}$.  $T_{ghosts}$ is constructed
out of the $b,c$ and $\tilde b, \tilde c$ ghosts.  The central charges
of these stress tensors are
\eqn\centcharges{\eqalign{
&c_{SL(2)} = {3(Q_5+2) \over Q_5} + {3\over 2} \cr
&c_{SU(2)} = {3(Q_5-2) \over Q_5} + {3\over 2} \cr
&c_{\bf X}= 6\cr
&c_{ghosts}= -15.}}
As in the bosonic problem  the total central charge $c_{worldsheet} =
c_{SL(2)} +c_{SU(2)} + c_{\bf X} +c_{ghosts}$ vanishes.

We can now fix a lightcone gauge $J^+= Q_5+2$.  The BRST charge then
has a term proportional to $\tilde Q_{BRST}= \oint c (J^- +
\psi^3\psi^-)+ \tilde c \psi^- $, and higher order terms.  Again, the
BRST cohomology of the full BRST charge is the same as that of $\tilde
Q_{BRST}$.  As above, the spacetime stress tensor is obtained by
twisting the stress tensor \stressws\
\eqn\stresst{T_{total}= T_{worldsheet} + \partial (J^3 + \psi^+\psi^-)}
such that $J^+$, $J^-$, $\psi^+$ and $\psi^-$ have conformal
dimensions $0$, $2$, $-{1\over 2}$, and ${3\over 2}$ respectively. 
As in the bosonic problem, the twisting changes the central charge of
the stress tensor \stresst\ to
\eqn\ctotal{c_{total}= 6Q_5}
in agreement with the exact result above.

The physical degrees of freedom living on the long string include the
bosonic $SU(2)_{Q_5-2}$, the three free fermions $\chi^a$, the modes
of the sigma model on ${\bf X}$, the $\phi$ boson and its superpartner
which we denote by $\psi$.  The other modes on $\AdS$ and the ghosts
are effectively removed by considering the cohomology.  The central
charge \ctotal\ is obtained because the boson $\phi$ becomes a
Liouville field with an improvement term $Q=(Q_5-1)\sqrt{2\over Q_5}$.
Because of this improvement term the system has a gap
\eqn\exactgap{\Delta_0= {Q^2\over 8} = {(Q_5-1)^2 \over 4Q_5}.}
This result is consistent with the semiclassical answer $Q_5 \over 4$
we found above.

Our identification of the degrees of freedom on the long string and
the fact that this theory is essentially a free CFT depend on having a
single long string; i.e.\ $m=1$ in \qgeneral.  For $m>1$ coincident
long strings, the collective coordinate $\phi$ describes their center
of mass, but there are also other degrees of freedom corresponding to
the separation between them.  These degrees of freedom lead to
$c=6Q_5m $ rather than $c=6Q_5$.  As is common in D-branes, the center
of mass theory can be free but the remaining degrees of freedom are
interacting.

\bigskip\noindent{\it Spacetime Supersymmetry}

Finally, we will present a description of the long string that is more
precise than the above and exhibits spacetime supersymmetry.  For
this, we will use a sort of unitary gauge description in terms of
physical degrees of freedom only, an approach we followed already in
section 3.  The reason for using unitary gauge is that to see
spacetime supersymmetry, one would like to use a Green-Schwarz type
description, but the Green-Schwarz string is difficult to quantize
covariantly\foot{For a recent attempt to derive a somewhat similar
description of $\AdS_3$ models from the Green-Schwarz string, see
\ref\pesando{I. Pesando, ``On The Quantization Of The GS Type IIB
Superstring Action On $\AdS_3\times \S^3$ With NS-NS Flux,''
hep-th/9903086.}.}. 

For simplicity, consider the case
${\bf X}= {\bf T}^4$ (the extension to ${\bf K3}$ is straightforward).
We start with an RNS description and then pass to Green-Schwarz
variables by introducing spin fields.  
The sigma model on ${\bf X}$ is described by four free bosons $x^i$
and four free fermions $\psi^i$.  Motion on
$\S^3$ is described by an $SU(2)$ current
algebra.  In a gauge $\gamma=z$, the long string motion on $\AdS_3$
is described by a Liouville field $\phi$ introduced in section 3.
In the RNS description, two worldsheet fermions that are superpartners
of $\gamma,\bar\gamma $ can be set to zero by a gauge choice.  In this
unitary gauge, all ghosts decouple.  So the description is by $x^i$, $\phi$,
and the $SU(2)$ current algebra, plus eight fermion partners.

To go to a Green-Schwarz description with manifest spacetime supersymmetry,
one replaces the eight RNS fermions by their eight spin fields, which
have dimension $1/2$ and are free fermions.  In terms of these
eight fermion fields along with $x^i$, $\phi$, and the $SU(2)$ currents
$j^a$, we want to construct an ${\cal N}=4$ superconformal algebra
which we will interpret as the spacetime superconformal algebra.

The $x^i$ together with four free fermions make an ${\cal N}=4$
hypermultiplet, with $c=6$.  So the essential point is to construct
{}from $\phi$, the $SU(2)$ currents $j^a$ at level $Q_5-2$, and four
free fermions $S^\mu$, an ${\cal N}=4$ algebra that, roughly,
describes the long string motion on $\AdS_3\times \S^3$.  The
nontrivial operator product expansions are
\eqn\nontriope{\eqalign{
&S^\mu(z)S^\nu(w) \sim -{\delta^{\mu\nu} \over z-w}\cr
&\partial\phi(z)\partial\phi(w) \sim  -{1 \over (z-w)^2} \cr
&j^a(z)j^b(w) \sim -{\delta^{ab} (Q_5-2)/2\over (z-w)^2} +
{\epsilon^{abc}j^c \over (z-w)}.\cr}}
We use the 'tHooft $\eta$
symbol 
\eqn\etac{\eta^a_{\mu\nu}= \half (\delta_{a\mu}\delta_{0\nu}-
\delta_{a\nu}\delta_{0\mu} +\epsilon_{a\mu\nu})}
to express the ${\cal N}=4$ generators
\eqn\generators{\eqalign{
&T=-\half \partial S^\mu S^\mu -{j^aj^a\over Q_5}
-\half\partial\phi\partial\phi +{\sqrt 2 (Q_5-1) \over 2\sqrt{Q_5}}
\partial^2\phi \cr  
&J^a=j^a+\half\eta^a_{\mu\nu}S^\mu S^\nu\cr
&G^\mu= {\sqrt 2 \over 2}\partial\phi S^\mu - {2 \over
\sqrt {Q_5}} \eta^a_{\mu\nu}j^aS^\nu +{1\over 6\sqrt {Q_5}}
\epsilon_{\mu\nu\rho\sigma}S^\nu S^\rho S^\sigma - {Q_5-1\over\sqrt
{Q_5}} \partial S^\mu .\cr}}
A straightforward calculation shows that they satisfy the
$\NN=4$ algebra with $c=6(Q_5-1)$
\eqn\nfour{\eqalign{
&J^a(z)J^b(w) \sim -{\delta^{ab} (Q_5-1)/2\over (z-w)^2} +
{\epsilon^{abc}J^c \over (z-w)}\cr
&T(z)J^a(w) \sim {J^a \over (z-w)^2} + {\partial J^a \over (z-w)}\cr
&T(z)T(w) \sim {3(Q_5-1) \over(z-w)^4} +{2T \over (z-w)^2} +
{\partial T\over z-w} \cr
&T(z)G^\mu(w) \sim {{3\over 2} G^\mu \over (z-w)^2} + {\partial G^\mu
\over (z-w)}\cr 
&J^a(z)G^\mu(w) \sim {\eta^a_{\mu\nu} \over z-w} G^\nu \cr
&G^\mu(z)G^{\nu}(w) \sim {\delta^{\mu\nu} 2(Q_5-1) \over
(z-w)^3} -{4 \over (z-w)^2} \eta^a_{\mu\nu} J^a + {1 \over z-w}
\left(\delta^{\mu\nu}T - 2 \eta^a_{\mu\nu} \partial
J^a\right).} }
Together with the  $\NN=4$ algebra of the ${\bf T}^4$, which has $c=6$,
 we have the
expected result of $\NN=4$ with $c=6Q_5$.

\nref\sch{K.~Schoutens, ``O(N) Extended Superconformal Field Theory In
Superspace,'' Nucl. Phys. {\bf B295} (1988) 634.}%
\nref\stvp{A.~Sevrin, W.~Troost and A.~Van Proeyen, ``Superconformal
Algebras In Two-Dimensions With N=4,'' Phys. Lett. {\bf 208B} (1988)
447.}%
\nref\fikl{E.A.~Ivanov, S.O.~Krivonos and V.M.~Leviant, ``A New Class
Of Superconformal Sigma Models With The Wess-Zumino Action,'' 
Nucl. Phys. {\bf B304} (1988) 601; ``Quantum N=3, N=4 Superconformal
WZW Sigma Models,'' Phys. Lett. {\bf B215} (1988) 689.}%
\nref\KPR{C.~Kounnas, M.~Porrati and B.~Rostand, ``ON N=4 extended
super-Liouville theory,'' Phys. Lett. {\bf B258} (1991) 61.}%

If we remove the improvement terms (last terms) in $T$ and $G^\mu$ and
add the free fermions $S^\mu$, the $U(1)$ current $\partial \phi$
and the $SU(2)_1$ currents $\tilde J^a=\half \bar \eta^a_{\mu\nu}S^\mu
S^\nu$, we find a realization of the large $\NN=4$ algebra
\refs{\sch-\KPR}.  This algebra has two different ordinary $\NN=4$
subalgebras: the one above with $c=6(Q_5-1)$, and another one
including $\tilde J^a$ but without $J^a$ with $c=6$.  The other
$\NN=4$ algebra appeared in the study of string propagation on
solitons
\ref\chs{C.G.~Callan, J.A.~Harvey and A.~Strominger, ``World Sheet
Approach to Heterotic Instantons and Solitons,'' Nucl. Phys. {\bf
B359} (1991) 611; ``Worldbrane Actions for String Solitons,''
Nucl. Phys. {\bf B367} (1991) 60; ``Supersymmetric String Solitons,''
hep-th/9112030.}.  

These two constructions are also important in the closely related
gauge theory on the onebranes.  The algebra with $c=6$ appears in the
tube of the Coulomb branch \seidai.  The algebra \generators\ with
$c=6(Q_5-1)$ appears along the Higgs branch of the same system.  Note
that as in \witcom, they have different R symmetries and different
central charges.  

The chiral operators of the $\NN=4$ algebra of the long string
theory \nfour\ are easily found to be
\eqn\chirops{\CO_j= e^{j\sqrt{2\over Q_5}\phi} V_j \qquad \qquad 
j=0,{1 \over 2},...,{Q_5-2\over 2} ,} 
where $V_j$ are the spin $j$
operators in the bosonic $SU(2)_{Q_5-2}$ WZW theory.  The exponent 
in the first factor is determined by imposing
\eqn\chicon{\Delta(\CO_j)=j,}
or by demanding that there is no isospin $j+\half$ operator in the
simple pole in the operator product expansion of $G^\mu \CO_j$.

The chiral operators $\CO_\half$ lead to the following four desendants
of conformal dimension $(1,1)$
(the other descendants at this level are null since $\CO_\half$ is
chiral):
\eqn\deltaL{\delta L=\{\bar G,[G,\CO_\half]\}.}
They are invariant under the left moving and right moving
$SU(2)_{Q_5-1}$ current algebras in the $(4,4)$ symmetry.  As is
standard in $\NN=4$ superconformal field theories, such operators are
truly marginal and preserve the $(4,4)$ superconformal symmetry.  With
these operators added to the Lagrangian we find $\NN=4$ Liouville.

As we explained in section 3, if one perturbs the system so that the long
string is not BPS, such a Liouville potential is generated.  Let
us check that the operators \deltaL\ have the expected quantum numbers.
The operators \deltaL\ transform as $({\bf 2},{\bf 2})$ under the
$SU(2)\times SU(2)$ outer automorphism of the $(4,4)$ superconformal
algebra.  In the theory of the Higgs branch -- understood in terms of
a small instanton on $\R^4 $ -- the outer automorphism  group acts by
rotations of $\R^4$.  Since the noncompact bosons describing motion on $\R^4$
cannot be decomposed as sums of left and right-movers,
 only a diagonal
subgroup $SU(2)_D$  of the outer automorphism group is a symmetry.  
(If one embeds the small instanton in $\bX={\bf T}^4$ or ${\bf K3}$, 
this explicitly
breaks $SU(2)_D$.  But for a sufficiently small instanton, the $SU(2)_D$ 
symmetry
is a good approximation, as the small instanton singularity does not ``see''
the compactification of $\R^4$ to $\bX$.) 
Under $SU(2)_D$, the four operators
\deltaL\ transform as ${\bf 1}\oplus {\bf 3}$.  

Now we compare this to the D1/D5 system.
As reviewed in section 2.2, this system can 
be deformed away from a singular point by turning on a $\theta$ angle 
or FI terms.  These transform as ${\bf 1}\oplus {\bf
3}$ of $SU(2)_D$, just the quantum numbers of the $(1,1)$ operators
\deltaL.

Liouville theory with improvement term proportional to $Q$ has
operators $e^{\alpha \phi}$ with dimensions $\Delta_\alpha= -\half
\alpha(\alpha-Q)$.  The values of $\alpha$ are constrained by 
\liouvillen
\eqn\alphacon{\eqalign{
&\alpha\le{Q\over 2} \qquad {\rm for} ~Q>0 \cr
&\alpha\ge -{Q\over 2} \qquad {\rm for} ~Q<0 .\cr}}
The string coupling depends on $\phi$ according to
\eqn\gstphi{g(\phi)=e^{Q\phi/2},}
and therefore the coupling is strong as $\phi \rightarrow +\infty$
($\phi \rightarrow -\infty$) for $Q$ positive (negative).  The wave
function associated with the operator $e^{\alpha\phi}$ is 
\eqn\vaveop{\psi(\phi)={1\over g(\phi)} e^{\alpha\phi}= 
e^{(\alpha - {Q\over 2})  \phi}.}
The condition \alphacon\ can be interpreted as the condition that the
wave function diverges at the weak coupling end \liouvillen.

In our problem we have two Liouville systems.  The first is the
worldsheet theory in the $\AdS_3$ background.  For large $\phi$, it
becomes a free theory with an improvement term with $Q_C=-\sqrt{2\over
Q_5}$.  This value is obtained by looking at the shift term of the
free Wakimoto description of our system \gks.  Alternatively, it can
also be obtained by analyzing the theory on the Coulomb branch and its
tube (hence the subscript $C$) \chs.  This system describes the short
strings.  The second Liouville system is on the long string and
corresponds to the tube of the Higgs branch.  From that point of view
we have seen that large $\phi$ corresponds to strong coupling.
The Liouville system of the long strings has
$Q_H=(Q_5-1)\sqrt{2\over Q_5}$.  $Q_C$ and $Q_H$ differ in sign and in
the absolute value.  These two facts 
follow from the two $N=4$ superconformal
subalgebras of the large $N=4$ \stvp.

It is natural to identify $\phi$ of the two tubes.  The string
coupling for the short strings is large at one end and the string
coupling for the long strings is large at the other end:
\eqn\strcouch{\eqalign{
&g_C(\phi)=e^{Q_c\phi/ 2}=\exp({- {1 \over \sqrt{ 2Q_5}} \phi}) \cr
&g_H(\phi)=e^{Q_H\phi/ 2}=\exp({{Q_5-1 \over \sqrt{ 2Q_5}}\phi}).\cr}}

The vertex operators $e^{\alpha \phi}$ can act either on short strings
or on long strings.  Therefore, they should make sense in the two
Liouville systems and hence\foot{Strictly, the bound on $\alpha$ in
$SL(2)$ could be weaker.  Here, we will use the bound in Liouville
theory, which is obtained by studying the tube of the Coulomb branch.}
\eqn\alphacondh{-{1\over \sqrt{2Q_5}} \le \alpha \le {Q_5-1 \over 
\sqrt{2 Q_5}}.}
This guarantees that their wave functions diverge at the weak coupling
end both for short strings and for long strings.  

The fact that the wave function $\psi(\phi)$ of a local vertex
operator ${\cal O}$ diverges at large $\phi$ for
a small string state but vanishes exponentially at large $\phi$ for
a long string seems strange at first, but is not so hard to
understand intuitively.  It means roughly that ${\cal O}$ is likely
to create a small string near the boundary of $\AdS_3$, but is very
unlikely to create a long string wrapped around the boundary.

It is reassuring to see that the chiral operators \chirops\ satisfy
\alphacondh.  We will have more remarks about this bound in section 5.

The chiral operators \chirops\ can be compared with their counterparts
in the first quantized description of \gks\ (equation (4.11)):
\eqn\fourele{\int d^2z e^{-\varphi-\bar\varphi}(\psi^3-\half
\gamma\psi^- -\half \gamma^{-1}\psi^+)(\bar \psi^3-\half\bar 
\gamma\bar\psi^- -\half \bar \gamma^{-1}\bar \psi^+) 
e^{j\sqrt{2\over Q_5}\phi} \gamma^{j+m}\bar\gamma^{j+\bar m}V_j,}
where $\varphi$ and $\bar \varphi$ are the bosonized ghosts.  This
expression is in the minus one ghost picture.  In the zero picture the
exponential of the ghosts disappears and the factors with the fermions
are replaced by the $SL(2)$ currents and some higher order terms.  We
follow the light-cone like gauge fixing \currc\ and keep only the term
proportional to $J^+$, which is a constant in our gauge.  Using
$\gamma=z$ we recognize \fourele\ as the $(m,\bar m)$ mode of the
local operator $e^{j\sqrt{2\over Q_5}\phi} V_j$.  This is the chiral
operator \chirops.

\bigskip\noindent{\it Remarks On $Q_5=1$}

The near horizon NS5/NS1 system does not seem to make sense for
$Q_5=1$, since it uses an $SU(2)$ current algebra at level 
$Q_5-2$ \seidai.  Hence in particular the above discussion
does not make sense for $Q_5=1$.  However, there is a sense in
which one can approach $Q_5=1$: as reviewed in section 2, we can define
a model that depends on an arbitrary positive integer $N$ and
move around in its moduli space until we reach the locus where the
``small instanton singularity of $Q_5=1$'' should appear.  In this
sense, something must happen as one approaches $Q_5=1$.

Though our derivation of \exactgap\ does not make sense for $Q_5=1$
(since the conformal field theory we used is not defined there), it is
tempting to believe that the formula is still valid for that value of
$Q_5$.  If so, the gap vanishes at $Q_5=1$.  We take this to mean that
the ground state of the spacetime CFT becomes unnormalizable when one
moves around in the parameter space \uxu\ and approaches the locus of
the $Q_5=1$ singularity.  This is compatible intuitively with the
argument at the end of section 2 indicating that the singularity is
worse for $Q_5=1$, but we have no detailed interpretation to offer.

One might guess that the $Q_5=1$ system simply describes the
scattering of low energy particles in spacetime from a single NS
fivebrane in the weak coupling limit.  If so, the essential difference
between $Q_5=1$ and $Q_5>1$ is that (as there is \seidai\ no tube for
$Q_5=1$) the conformal field theory of a single fivebrane (unlike that
of $Q_5>1$ coincident fivebranes) is nonsingular, and hence string
perturbation theory is well behaved as the string coupling goes to
zero.  The weak coupling limit describes particles in spacetime that
scatter from a potential (the fivebrane) but not from each other.
Perhaps as one approaches the $Q_5=1$ singularity, the near horizon
CFT somehow describes this physics.  Hence, one may wonder whether the
$Q_5=1$ system is related to the symmetric product $(\R^4\times
\bX)^{Q_1}/S_{Q_1}$.  Clearly, this issue deserves better
understanding.

\newsec{Missing Chiral States, And Further Applications}

\subsec{The Fate Of Chiral Primaries}

\nref\vafastrom{A. Strominger and C. Vafa, ``Microscopic Origin Of The
Bekenstein-Hawking Entropy,'' hep-th/9601029, Phys. Lett. {\bf B379}
(1996) 99.}  
The analysis of the worldsheet conformal field theory of the NS1/NS5
theory in \gks\ has found chiral operators in the boundary conformal
field theory with $\Delta \le {Q_5-2\over 2}$.  However, general
considerations from a spacetime point of view \vafastrom\ suggest that
the system has chiral operators with $\Delta\le {Q_1Q_5\over 2}$.
Where are the missing operators?  The analysis of \vafastrom\ is valid
for generic values of the parameters, where the 
CFT is not singular.  On the other hand, the analysis of \gks\ is
valid on a sixteen-dimensional subspace of the system, where all the
RR moduli of the NS1/NS5 system vanish.  Precisely on this subspace,
the system has, as we have seen, a continuum starting at
$\Delta_0={(Q_5-1)^2 \over 4Q_5}$.  It is natural to suspect
that the disappearance of some states when the RR moduli vanish
are somehow associated with the appearance of this continuum or
(shifting to a D1/D5 language) with the small instanton singularity.

\bigskip\noindent{\it Comparison To Classical Instanton Moduli Space}

To get some insight about this, let us look at the small instanton
singularity classically.  According to the ADHM construction, the
moduli space of instantons on $\R^4$ is the Higgs branch of the
following theory with eight supercharges: a $U(1)$ gauge theory with
$Q_5$ hypermultiplets ${\cal A}^i$, $i=1,\dots, Q_5$ of charge 1.
{}From the point of view of a theory with only four supercharges, each
${\cal A}^i$ splits as a pair of chiral multiplets $A^i$, $B_i$.  The
equations determining a vacuum are $F$-flatness
\eqn\bumblebee{\sum_{i=1}^{Q_5} A^iB_i=0}
and $D$-flatness
\eqn\numblebee{\sum_{i=1}^{Q_5}|A^i|^2 -\sum_{j=1}^{Q_5}|B_j|^2=r.}
Here we have included an FI interaction with coefficient $r$.  For
$r\not= 0$, the small instanton singularity is ``blown up''; the singularity
 is recovered
for $r=0$.  Up to an R-symmetry transformation (rotating the choice of
four supercharges out of the eight), there is no loss of generality
in assuming that the FI term appears only in \numblebee\ and that $r>0$.

The small instanton moduli space ${\cal T}$ is the space of solutions of the 
above
equations, modulo the $U(1)$ action $A\to e^{iw}A$, $B\to e^{-iw}B$.  
Setting $B=0$, we see that ${\cal T}$ contains a copy of $U={\bf CP}^{Q_5-1}$,
of radius $\sqrt r$.  ${\cal T}$ itself is the cotangent bundle $T^*U=
T^*{\bf CP}^{Q_5-1}$
of $U$, as long as $r>0$.  For $r=0$, the copy of ${\bf CP}^{Q_5-1}$
at the ``center'' of ${\cal T}$ is blown down to a point, and ${\cal T}$
has a conical singularity.

Now for topological reasons, for $r>0$, there is an ${\bf L}^2$ harmonic form
of the middle dimension on ${\cal T}$.  Indeed,
a delta function that is Poincar\'e dual to $U$ is
of compact support and also (since the fibration $T^*U\to U$ 
has nonzero Euler class) represents a nontrivial cohomology class.
It can therefore be projected to a unique nonzero ${\bf L}^2$ form $\omega$ on 
${\cal T}$.
For $r>0$, $\omega$ generates the image of the compactly supported cohomology
of ${\cal T} $ in the ordinary cohomology of this space,
and hence generates the ${\bf L}^2$ cohomology of ${\cal T}$.

Being
Poincar\'e dual to a complex submanifold of complex codimension $Q_5-1$,
$\omega$ is a form of degree $(Q_5-1,Q_5-1)$ on ${\cal T}$.  It
hence corresponds, for $r\not= 0$, to a chiral primary field
of dimension $((Q_5-1)/2,(Q_5-1)/2)$ -- exactly the
quantum numbers of the first chiral primary that is ``missing''
in the conformal field theory analysis of the NS1/NS5 system without
RR fields \gks.

The ${\bf L}^2$ form $\omega$ has its support for $A,$ $B$ of order
$\sqrt r$, since this is the radius of $U$.  
If, therefore, we approach the locus on which some chiral
primaries are ``missing'' by taking $r\to 0$, then $\omega$ has
delta function support at the singularity of the moduli space.
(We use the phrase ``delta function support'' somewhat loosely to 
mean simply that as $r\to 0$, $\omega$ vanishes away from the conical 
singularity.)
A more evocative way to say this is that as $r\to 0$, $\omega$ becomes
concentrated at the singularity and disappears
from the smooth part of ${\cal T}$.

Let us now try to interpret this in the conformal field theory.  The
fields $A$, $B$ do not give a good description of the small instanton
conformal field theory near the singularity.  For this we should use
instead the Liouville field $\phi$ of the long strings together with other
degrees of freedom described in section 4.  The small
instanton region corresponds in that description to $\phi\to\infty$.
For $r$ very small and positive, an exponential coupling $e^\phi$ suppresses
the region of large $\phi$.  The fact that the classical form $\omega$
has its support at $A,B\sim \sqrt r$ suggests that the chiral primary
corresponding to $\omega$ in the (better) long string description has
its support at, roughly,
$e^\phi \sim 1/r^\delta $ for some $\delta>0$.  For $r\to 0$,
this chiral primary state then disappears to $\phi=\infty$.

To prove that a chiral primary state corresponding to $\omega$ moves
to $\phi=\infty$ as $r\to 0$, we would need a fuller understanding of
the ${\cal N}=4$ Liouville theory.  But this behavior is certainly
suggested by the classical facts that we have explained, 
and is entirely consistent with the
fact that for $r=0$, that is in the absence of RR fields, a chiral
primary with the same quantum numbers as $\omega$ is missing in the
NS1/NS5 conformal field theory.

The discussion so far makes it sound as if only one state should be 
``missing,''
but that is because we have focussed on just a single small instanton.
Let us see what happens when we incorporate the other degrees of freedom.
We let ${\cal M}_{Q_1,Q_5}$ be the moduli space of $U(Q_5)$ instantons of 
instanton
number $Q_1$.  The locus of ${\cal M}_{Q_1,Q_5}$ parametrizing configurations
with a very small instanton looks
like a product (more precisely, a fiber bundle) ${\cal T}\times \bX\times
{\cal M}_{Q_1-1,Q_5}$, where ${\cal M}_{Q_1-1,Q_5}$ parametrizes
the $Q_1-1$ non-small instantons, $\bX$ parametrizes the position of
the small instanton, and ${\cal T}$ is the one-instanton moduli space
discussed above.  Let $\alpha$ be any harmonic form
 on $\bX\times {\cal M}_{Q_1-1,Q_5}$.  Then, when ${\cal T} $ is blown
down by turning off the RR fields, $\omega\wedge \alpha$ has its support
at the small instanton singularity in ${\cal T}$ (since $\omega$ does),
so every cohomology class of this form disappears from the smooth part of
the moduli space in this limit.  

Harmonic forms on ${\cal M}_{Q_1,Q_5}$ of degree $(p,q)$ with
$p$ (or $q$) less than $Q_5-1$ are supported, for $r\to 0$, on the smooth
part of the moduli space (since the singular part does not support any
${\bf L}^2$ harmonic form of such degree).  Hence, the chiral primaries
with $j<(Q_5-1)/2$ should decay as one approaches the long string or
small instanton region, as we found in section 4.

\subsec{Further Applications}

We conclude by briefly remarking on some further applications.

In the NS1/NS5 case, the essence of our result concerns the behavior
of long strings in the WZW model of $SL(2,{\bf R})$.  
This model has been studied over the years from many point of view, 
for instance as an example of a non-unitary CFT (see,
e.g.
\nref\zamfat{A.B.~Zamolodchikov and V.A.~Fateev,
``Operator Algebra And Correlation Functions In The Two-Dimensional
Wess-Zumino $SU(2)\times SU(2)$ Chiral Model,''
Sov. J. Nucl.~Phys. {\bf 43} (1986) 657.}\nref\tesch{J. Teschner, ``The
Minisuperspace Limit Of The $SL(2,{\bf C})/SU(2)$ WZW Model, hep-th/9712258.}%
\refs{\zamfat,\tesch}),
an example of string propagation in nontrivial and time-dependent
backgrounds (see, e.g.\
\nref\dlp{L.J.~Dixon, M.E.~Peskin and J.~Lykken, ``N=2 Superconformal
Symmetry And SO(2,1) Current Algebra,'' Nucl. Phys. {\bf B325} (1989)
329.}%
\nref\egp{I. Bars and D. Nemeschansky, ``String Propagation In Backgrounds
With Curved Space-Time,'' Nucl. Phys. {\bf B348} (1991) 89.}%
\refs{\perry,\dlp,\egp} and references therein),  as an ingredient in
studying two-dimensional quantum
gravity 
\ref\kpz{V.G.~Knizhnik, A.M.~Polyakov and A.B.~Zamolodchikov,
``Fractal Structure Of 2-D Quantum Gravity,'' Mod. Phys. Lett. {\bf
A3} (1988) 819.},  
and black holes
\ref\wittenbh{E.~Witten, ``On String Theory and Black Holes,"
Phys. Rev. {\bf D44} (1991) 314.}, 
and in the context of the AdS/CFT correspondence \refs{\gks - \ks}.
This much-studied
two-dimensional CFT has the strange behavior we have described: a
continuum (above a certain threshold) coming from long strings, above
and beyond the continuum coming from the noncompactness of the group
manifold.  This raises the interesting question, which we will not try
to address, of what sort of singularities in correlation functions are
generated by the long strings.

\bigskip\noindent
{\it Breakdown Of String Perturbation Theory}

\nref\huto{C.M.~Hull and P.K.~Townsend, ``Unity of Superstring
Dualities,'' Nucl. Phys. {\bf B438} (1995) 109, hep-th/9410167.}%
\nref\wittenvarious{E.~Witten, ``String Theory Dynamics in Various
Dimensions,'' Nucl. Phys. {\bf B443} (1995) 85,hep-th/9503124.}%
In certain applications of conformal field theories of the sort we have
been studying, the behavior we have found has dramatic consequences.
The Type IIA string on ${\bf K3}$ is dual to the heterotic string
on $\T^4$ \refs{\huto,\wittenvarious}; using this duality one can show that
when Type IIA string theory is compactified on a space $M$ with an
$A_N$ singularity, an enhanced $A_N$ gauge symmetry occurs in the six
dimensions orthogonal to $M$.  This is a nonperturbative phenomenon
that cannot be avoided by making the string coupling constant smaller.
It thus represents a breakdown of string perturbation theory,  possible
only because of a singular behavior of the conformal field theory. 
This breakdown of perturbation theory is associated with an important
role played by wrapped membranes.  One might suspect that the
breakdown of perturbation theory is associated with the appearance
of a ``tube'' with a linearly growing coupling.
  However, the
target space of the CFT does not exhibit such a singularity; indeed,
the hyper-K\"ahler metric of the target space is subject to no quantum 
corrections.  In fact,
in some cases the target space has only an orbifold singularity and
the theory differs from the non-singular orbifold only in the
value of the $\theta$ angle
\ref\aspinwall{P.S.~Aspinwall, ``Enhanced Gauge Symmetries and K3
Surfaces,'' Phys. Lett. {\bf B357} (1995) 329, hep-th/9507012.}.

Despite this, for string theory at the $A_{n-1}$ singularity,
a tubelike description arises in suitable variables
via $T$-duality \refs{\vafoog,\ghm}. 
These examples are similar to the ones we have been studying since,
for example, the $A_1$ singularity is the small instanton singularity
for $Q_5=2$.  For the D1/D5 cases that we have studied in the present
paper, we have seen a mechanism
by which  the CFT  develops
the expected tube in field space (but not in the original variables
of the sigma model), and therefore the CFT is singular.  Thus, our result
can be seen as an analog or generalization of the tubelike description
of the $A_{n-1}$ singularities that comes from $T$-duality.

The D0/D4 system compactified on $\bf X$ is closely related to the
D1/D5 system.  Its low energy behavior is described by quantum
mechanics whose target space is the moduli space of instantons on $\bf
X$.  For certain values of the parameters (one must set the three FI couplings
to zero), this
target space has singularities.  They are associated either with small
instantons or reduced structure group.  The pathology discussed
in this paper is a feature of $1+1$-dimensional conformal field theory
and does not affect D0/D4 quantum mechanics, which remains
well behaved even with the target space develops singularities.
However, some low energy states in the quantum mechanics are supported
near the singularities, for reasons explained in section 5.1.
In making this assertion, we are assuming that, when one turns off
the  FI couplings, the states in question do not spread on
the Coulomb branch.  Assuming this, these
states are a generalization, to higher $Q_5$, of the D0/D4 bound
state at threshold whose existence was proved in
\ref\sethi{S. Sethi and M. Stern, ``A Comment On The Spectrum Of
$H$-Monopoles,'' hep-th/9607145.}.
\bigskip
\centerline{\bf Acknowledgements}

We would like to thank O. Aharony, T. Banks, R. Dijkgraaf,
R. Friedman, D. Kutasov,
J. Maldacena, G. Moore, A. Strominger, and S.-T. Yau for useful
discussions.  N.S. thanks the organizers of the String Workshop at the
Institute for Advanced Studies at the Hebrew University for
hospitality and for creating a stimulating environment during the
completion of this work.  This work was supported in part by grant
\#DE-FG02-90ER40542 and grant \#NSF-PHY-9513835.

\listrefs

\end